\documentclass{article}
\usepackage[utf8]{inputenc}
\usepackage{amsmath}
\usepackage{amsfonts}
\usepackage{fullpage}
\usepackage[dvipsnames]{xcolor}
\usepackage{appendix}
\usepackage{biblatex}
\usepackage{multirow}
\usepackage{graphicx}
\usepackage{changepage}

\bibliography{biblio}

\usepackage{soul}

\def\boxit#1{\vbox{\hrule\hbox{\vrule\kern6pt
          \vbox{\kern6pt#1\kern6pt}\kern6pt\vrule}\hrule}}

\title{Spatial Autoregressive Model on a Dirichlet Distribution}
\author{Teo Nguyen, Sarat Moka, Kerrie Mengersen, and Benoit Liquet}
\date{}

\begin{document}

\maketitle

\section*{Abstract}

Compositional data find broad application across diverse fields, including ecology, geology, economics, and public health, due to their efficacy in representing proportions or percentages of various components within a whole. Spatial dependencies often exist in compositional data, particularly when the data represents different land uses or ecological variables. Spatial autocorrelation can arise from shared environmental conditions or geographical proximity, and ignoring these autocorrelations in modelling of compositional data may lead to incorrect estimates of parameters. Therefore, it is essential to incorporate spatial information into the statistical analysis of compositional data to obtain accurate and reliable results. However, traditional statistical methods are not directly applicable to compositional data due to the correlation between its observations, which are constrained to lie on a simplex. To address this challenge, the Dirichlet distribution is commonly employed, as its support aligns with the nature of compositional vectors. Specifically, the R package {\sf DirichletReg} provides a regression model, termed Dirichlet regression, tailored for compositional data. However, this model fails to account for spatial dependencies, thereby restricting its utility in spatial contexts.
In this study, we introduce a novel spatial autoregressive Dirichlet regression model for compositional data, adeptly integrating spatial dependencies among observations. We construct a maximum likelihood estimator for a Dirichlet density function augmented with a spatial lag term. 
We compare this spatial autoregressive model with the same model without spatial lag, where we test both models on synthetic data as well as two real datasets, using metrics such as $R^2$, RMSE, cosine similarity, cross-entropy or AIC. By considering the spatial relationships among observations, our model provides more accurate and reliable results for the analysis of compositional data. The model is further evaluated against a spatial multinomial regression model for compositional data, and their relative effectiveness is discussed.

\paragraph{Keywords} Compositional Data; Dirichlet Regression; Multinomial Regression; Spatial Autoregressive Model.

\section{Introduction}

Compositional data are widely used across diverse fields including ecology \cite{vercelloni2020forecasting}, chemistry \cite{egozcue2024exploring, ullah2023comprehensive}, economics \cite{thomas2021spatial,vo2023socio}, and public health \cite{blodgett2023associations,voinier2023association} owing to their capacity to carry relative information and to be represented as proportions or percentages \cite{aitchison1982statistical}. 

Mathematically, a $D$-part compositional dataset is defined as a vector $y = (y_1, y_2, \dots, y_D)\in \mathbb R^D$ such that,
\begin{equation*}
\left\{ 
\begin{array}{l} 
y_i \geq 0, \quad \forall \, i \in \{1,2,\dots,D\},\\
\displaystyle \sum_{i=1}^D y_i = 1.  \end{array}\right.
\end{equation*}
A simplex $S^D$ is defined as the set of all the $D$-part compositional data, i.e.
\begin{equation*}
    S^D = \Big\{ y = (y_1, y_2, ..., y_D) \ |\ \forall i \in \{1,2,\dots,D\}, y_i \geq 0 ; \sum_{i=1}^D y_i = 1  \Big\}.
\end{equation*}
Despite their widespread utilization, compositional data pose a distinctive challenge to statistical analysis due to their inherent relative nature and the constraint of lying on a simplex \cite{aitchison1982statistical}. Thus, traditional statistical methods cannot be applied directly to these types of data. To address this challenge, it is common in practice to assume that the probability distribution for compositional data is a Dirichlet distribution of parameter $\alpha \in \mathbb R^D$; a definition provided later in the next section. 
Maier in 2014 \cite{maier2014dirichletreg} proposed a regression model for compositional data under the assumption that the data are generated from a Dirichlet distribution, and this model is implemented in the R package {\sf DirichletReg}. However, unfortunately, {\sf DirichletReg} model does not take spatial dependencies into account, limiting its applicability in spatial problems.

Over the past few decades, spatial autoregressive (SAR) models have emerged as powerful tools for analyzing spatially correlated data in various fields mentioned above, including economics \cite{baltagi2008econometric}, ecology \cite{dormann2007methods}, and epidemiology \cite{hafner2020spread,raymundo2021spatial}. The fundamental idea behind SAR models is that the value of a variable at a particular location is influenced not only by its own characteristics but also by the characteristics of its neighboring locations. These models explicitly account for the spatial interdependencies among the observed variables, allowing for a more comprehensive understanding of the underlying spatial processes. While spatial dependencies are often present in compositional data, particularly when the observations represent different land uses or ecological variables, only a few studies have developed a SAR model for such data. Instead, in such studies, the authors employed either a Bayesian estimation approach to estimate the parameters of a spatial multinomial logit model \cite{krisztin2022spatial}, or transform the data into the Euclidian space before applying a multivariate regression model \cite{nguyen2021simultaneous}, a Gaussian Markov random field \cite{pirzamanbein2018modelling}, or a multivariate conditionally autoregressive model \cite{leininger2013spatial}. 

In order to incorporate spatial dependencies between observations in the case of compositional data, our paper proposes a SAR model. In particular, we develop a maximum likelihood estimator that effectively handles the spatial inter-dependencies between the observations by incorporating a spatial lag term into the regressor. This spatial lag term is defined through a correlation strength parameter $\rho \in [-1,1]$ and an adjacency matrix $W$ that contains the spatial information. This model will be referred as spatial lag model.

We demonstrate the effectiveness of our model on one synthetic dataset generated from a Dirichlet distribution, as well as three real-world datasets: the composition of sediments in a lake, the classification of corals within a lagoon, and the arrangement of votes during an election. Using different metrics such as $R^2$, Root Mean Squared Error, cross-entropy, and cosine similarity, the spatial lag Dirichlet model is compared to the Dirichlet model without spatial lag. We observe that the spatial lag model always perform better, excepted in the cases where no new information is added by the adjacency matrix.
The performances are also compared with the ones of a multinomial model, which is also particularly suited for modelling compositional data. The results show that, depending on the dataset, one model can perform better than the other. This specifically holds true when we compare the performances on a synthetic dataset generated via a Dirichlet distribution, and a synthetic dataset generated via a multinomial distribution.

The remaining paper is organized as follows.  The different models are described in Section \ref{sec:methodo}. Then, in Section \ref{sec:results}, the performances of the models are assessed through synthetic data and three case studies. A discussion is presented in Section \ref{sec:discussion}. Finally, we conclude the paper in Section \ref{sec:conclusion}.

\section{Materials and Methods}
\label{sec:methodo}

In this section, we first give a quick description of the Dirichlet distribution and its parameters. Then, the maximum likelihood regressors are presented, both without and with spatial lag. After that, the parallel between a multinomial regressor and the Dirichlet model is describe. Finally, the metrics used to analyse the results are presented.

We consider the case where the labels of the dataset are compositional. In this section and the following, let $K$ be the number of features, $J$ the number of classes, $n$ the sample size of the dataset. For the sample~$i$, we define its features as $x_i \in \mathbb R^K$ and its label $y_i \in S^J$ (i.e., $y_i$ is an element of the simplex of dimension $J$). The features (resp., labels) of the whole dataset are then denoted by $X \in \mathbb R^{n \times K}$ (resp., $Y \in \mathbb R^{n \times J}$).  If for a given data row $i$, the label $y_i$ follows a Dirichlet distribution of parameter $\alpha_i \in \mathbb R^J$, then the probability density function is
\begin{equation*}
f(y_i |  \alpha_i) = \frac{\Gamma (\sum_{j=1}^{J} \alpha_{ij})}{\prod_{j=1}^J \Gamma (\alpha_{ij}) } \prod_{j=1}^J y_{ij}^{\alpha_{ij}-1},
\end{equation*}
where $\Gamma$ is the gamma function, $\alpha_i = (\alpha_{i1}, \alpha_{i2}, \dots, \alpha_{iJ})$ is called the concentration parameter vector and has to meet the requirement that $\alpha_{ij} > 0$ for every class $j$.

The parameters $\alpha_i$ can be further parametrized by  $\alpha_i = \phi_i \mu_i$ where scalar $\phi_i \in \mathbb R$ is known as the precision parameter (or, dispersion parameter) and the compositional vector $\mu_i \in S^J$ represents the individual expected values. 
Then, all the parameters $\alpha_i$ can be stacked into a matrix $\alpha \in \mathbb R^{n \times J}$ such that $\alpha_i$ is the $i$th row of $\alpha$. Similarly, we can stack the $\mu_i$ (respectively, $\phi_i$) in a matrix $\mu \in \mathbb R^{n \times J}$ (respectively, in a vector $\phi \in \mathbb R^n$).

\begin{figure}
    \centering
    \includegraphics[width=14cm]{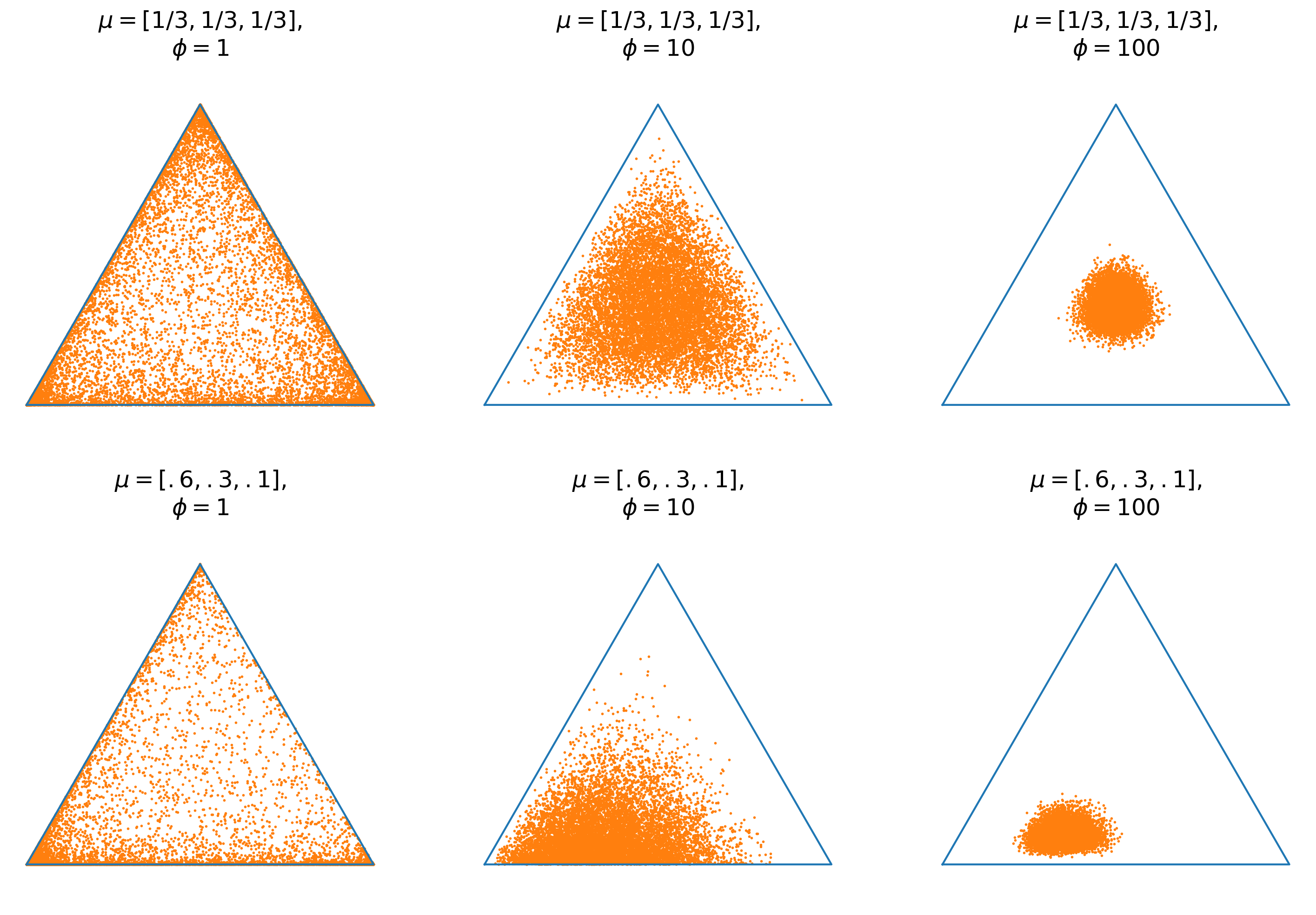}
    \caption{Scatter plot of 10000 sample points drawn from a Dirichlet distribution, for different values of $\mu \in S^3$ and $\phi \in \mathbb R$.}
    \label{fig:dirichlet_phi}
\end{figure}

The dispersion parameter $\phi_i$ plays a crucial role in distinguishing between the classes. For a given $\mu_i$, the smaller the value of $\phi_i$, the more likely the point will be distributed around extreme values (the edges of the simplex). On the contrary, with a high $\phi_i$, the point is more likely distributed close to the value of $\mu_i$. This effect is displayed in Figure \ref{fig:dirichlet_phi}, where 10000 sample points  were drawn from a Dirichlet distribution with different parameters $\phi$ and $\mu$.

Note that irrespective of the value of $\phi_i$, the expectation of these sample points is unchanged and remains $\mu_i$. However, in the case of our regression model, each point is drawn from a Dirichlet with different parameters (each observation has its own $\phi_i$). Hence, here, a high value of $\phi_i$ is preferred because the obtained vector will be close to $\mu_i$.

The estimated value of $\hat\phi_i$ may also be a good indicator of how accurate we expect the predicted value $\hat \mu_i$ to be: a large $\hat\phi_i$ should imply a more accurate $\hat\mu_i$.

\subsection{Maximum likelihood regression without spatial lag}

In this subsection, we introduce the same Dirichlet regression model as the one proposed by Maier \cite{maier2014dirichletreg}. Note that this method doesn't take spatial dependencies into consideration.

Let $\beta \in \mathbb R^{K \times J}$ be the matrix of coefficients. We define $\mu \in \mathbb R^{n \times J}$ as a matrix with elements,
\begin{equation}
\mu_{ij} = \frac{\exp( \sum_{k=1}^K X_{ik}\beta_{kj})}{\sum_{j'=1}^J \exp( \sum_{k=1}^K X_{ik}\beta_{kj'})}, \qquad i \in [1,\dots,n],\,\, j \in [1,\dots,J]. \label{eq:def_mu_non_spatial}
\end{equation}
Then, let $K_Z \in \mathbb N$ where $\mathbb N$ denotes the set of nonnegative integers. We introduce a matrix $Z \in \mathbb R^{n \times K_Z}$ and a vector $\gamma \in \mathbb R^{K_Z}$, that allow us to define the precision parameter vector $\phi \in \mathbb R^n$ with elements,
\begin{equation*}
\phi_i = \exp([Z\gamma]_i), \qquad i \in [1,\dots,n],
\end{equation*}
where the notation $[Z\gamma]_i$ corresponds to the $i$th row of the matrix $Z\gamma$.

For any row $i$ and class $j$, we then set $\alpha_{ij} = \phi_i \mu_{ij}$, so that $\phi_i = \sum_j \alpha_{ij}$. This parametrization is called the \emph{alternative} parametrization in Maier's paper \cite{maier2014dirichletreg}, contrasting it with the \emph{common} parametrization where each $\phi_i$ is set to 1.

To ensure the uniqueness of the solution when maximizing the likelihood, the mapping $\beta \mapsto \mu$ must be injective. Because of that, we have to set a column of $\beta$ as 0, for instance the first column, as done in \cite{maier2014dirichletreg}. The density function can be re-expressed as follows to emphasise the dependency on $\mu$ and $\phi$,
\begin{equation}
f(y_i | \mu_i, \phi_i) = \frac{\Gamma (\phi_i)}{\prod_{j=1}^J \Gamma (\phi_i \mu_{ij}) } \prod_{j=1}^J y_{ij}^{\phi_i \mu_{ij}-1}.
\label{eq:dirichlet_pdf}
\end{equation}
Thus, the log-likelihood of the Dirichlet distribution is,
\begin{align}
    \ell(y | \mu, \phi) &= \sum_{i=1}^n \left(
    \ln \Gamma ( \phi_i )
    - \sum_{j=1}^J \ln ( \Gamma(\phi_i \mu_{ij}) )
    + \sum_{j=1}^J ((\phi_i \mu_{ij}-1) \ln (y_{ij}))
    \right) \label{eq:loglikelihood} \\
    &= \sum_{i=1}^n \left(
    \ln \Gamma ( \phi_i )
    - \sum_{j=1}^J \ln \left( \Gamma\big(\phi_i \frac{\exp( [X\beta]_{ij} )}{\sum_{j'=1}^J \exp( [X\beta]_{ij'} \big) }) \right)
    + \sum_{j=1}^J \left((\phi_i \frac{\exp( [X\beta]_{ij})}{\sum_{j'=1}^J \exp( [X\beta]_{ij'})} - 1\right) \ln (y_{ij}))
    \right). \label{eq:loglikelihood_with_X}
\end{align}
The presence of the $\ln (y_{ij})$ term in this likelihood function requires $y_{ij}$ to be strictly positive for all $i$ and $j$. To address the issue of zero values in the data, a possible transformation is to use $y^* = \frac{y(n-1) + 1/J}{n}$ as in \cite{maier2014dirichletreg}, and it ensures that the transformed values are positive and has the property that $\lim_{n \to +\infty} y^* = y$. For the rest of this paper, we still denote the data labels as $y$ and without loss of generality assume it does not contain any zero values, but note that the transformation can be applied if necessary.

Because $\mu$ and $\phi$ are parameterized by $\beta$ and $\gamma$, maximum likelihood estimators $\hat \beta$ and $\hat \gamma$ are used to estimate the parameters $\beta$ and $\gamma$, respectively. Further, to perform second order optimization and also  to obtain the covariance matrix, the gradient and hessian matrix are computed. The details of the computation can be found in Appendix \ref{appendix:without_spatial}. Let $\psi$ be the digamma function, i.e., the derivative of the logarithm of the gamma function. That is,
\begin{equation}
    \label{eq:digamma}
    \psi(x) = \frac{\partial}{\partial x} \ln \Gamma(x) = \frac{\Gamma'(x)}{\Gamma(x)}, \quad  x \in \mathbb R.
\end{equation}
For all $(p,d) \in [1,\dots,K]\times [1,\dots,J]$, we have,
\begin{equation}
    \frac{\partial}{\partial \beta_{pd}} \ell (y | \mu, \phi) = \sum_{i=1}^n \left( \phi_i X_{ip} \mu_{id} \bigg( \sum_{j=1}^J \mu_{ij} \big( \psi(\phi_i \mu_{ij}) - \ln(y_{ij}) \big) 
    - \psi(\phi_i \mu_{id}) + \ln(y_{id})
    \bigg)  \right), \label{eq:gradient}
\end{equation}
and for $k \in [1,\dots,K_Z]$,
\begin{equation}
    \frac{\partial}{\partial \gamma_k} \ell (y | \mu, \phi) = \sum_{i=1}^n Z_{ik} \phi_i \bigg(  \psi(\phi_i) + \sum_{j=1}^J \mu_{ij} \big( \ln (y_{ij}) - \psi(\phi_i \mu_{ij}) \big)  \bigg). \label{eq:derivative_gamma}
\end{equation}
To keep the presentation concise, the expressions of the Hessian matrix are provided only in Appendix~\ref{appendix:without_spatial}.

By using the Broyden–Fletcher–Goldfarb–Shanno (BFGS) algorithm for maximization of the log-likelihood, we estimate the parameters $\hat \beta$ and $\hat \gamma$.  Then, we are able to predict the label of an unseen data point $\tilde x \in \mathbb R^K$. This prediction is the compositional vector $\tilde \mu \in S^J$, computed from \eqref{eq:def_mu_non_spatial}. The probability vector $\tilde \mu$ is considered as being the predicted value of the label.

\subsection{Maximum likelihood regression with spatial lag}

In this
subsection, we present our novel Dirichlet regression model that takes spatial dependencies in to consideration.

To take into account spatial effects in the model, we introduce a {\em spatial lag} through the matrix $M = I_n - \rho W$, where $I_n$ is the identity matrix of size $n \times n$, the parameter $\rho \in \mathbb R$ is the strength of spatial correlation, and $W \in \mathbb R^{n \times n}$ is the spatial weights matrix \cite{anselin1988spatial}, with $W_{pq}>0$ for all $p,q \in [1,\dots,n]$.  
It is common to apply row-normalization on $W$ (i.e., sum of each row is 1) \cite{anselin1988model}, which can often make it asymmetric even if the original non-normalized weights matrix is symmetric. 

Given a matrix of coefficients $\beta \in \mathbb R^{K \times J}$, we take $E_i^\Sigma = \sum_{j'=1}^J \exp( \sum_{i'=1}^n \sum_{k=K}^n M_{ii'}^{-1} X_{i'k} \beta_{kj'}) $ and redefine $\mu$ as a matrix with elements,
\begin{equation}
    \mu_{ij} = \frac{\exp([M^{-1} X \beta]_{ij})}{\sum_{j'=1}^J \exp([M^{-1} X \beta]_{ij'})} = \frac{\exp( \sum_{i'=1}^n \sum_{k=1}^K M_{ii'}^{-1} X_{i'k} \beta_{kj})}{E_i^\Sigma}. \label{eq:def_mu_spatial}
\end{equation}
The introduction of the matrix $M$, which allows to take spatial lag into consideration, modifies the computation of the vector $\mu$. In particular, multiplying the product $X\beta$ with the inverse of $M$ allows us to introduce the explanatory variables of the neighboring observations. The value of the spatial correlation parameter $\rho$ needs to be estimated from the data, while $W$ is fixed and has to be defined beforehand. Common choices for $W$ include distance-based weights or contiguity-based weights \cite{cliff1970spatial, o2003geographic}. 
In distance-based weights, the weight between each pair of points is determined by the inverse of the distance between them. In other words, points that are closer to each other receive higher weights, while points that are farther apart receive lower weights. In contiguity-based weights, for each points, the same weight is given to each of its nearest neighbours. The exact criteria for defining neighbors varies depending on the specific application and context, but it generally involves defining adjacency based on some spatial relationship between points.

Let us introduce the matrix $\tilde X \in \mathbb R^{n \times K}$ defined as $\tilde X = M^{-1}X$. With this notation, the loglikelihood remains the same as in~\eqref{eq:loglikelihood_with_X}, provided that we replace the term $X$ with $\tilde X$ in its expression. Similarly, the calculations of all the gradient and Hessian terms (w.r.t. $\beta$ and $\gamma$) of the loglikelihood are exactly similar to the ones without spatial lag, again providing that we replace $X$ by $\tilde X$. More specifically, we just have to replace $X_{ip}$ in \eqref{eq:gradient} by $\tilde X_{ip} = \sum_{i'} M^{-1}_{ii'} X_{i'p}$. Thus,
\begin{equation}
    \frac{\partial}{\partial \beta_{pd}} \ell (y | \mu, \phi, \rho) = \sum_{i=1}^n \left( \phi_i \tilde X_{ip} \mu_{id} \bigg( \sum_{j=1}^J \mu_{ij} \big( \psi(\phi_i \mu_{ij}) - \ln(y_{ij}) \big) 
    - \psi(\phi_i \mu_{id}) + \ln(y_{id})
    \bigg) \right). \label{eq:gradient_spatial}
\end{equation}
Unlike the model in the previous subsection, for this model with spatial lag, we also need to compute the derivative with respect to $\rho$. Towards this, by defining $U = M^{-1} W M^{-1} X \beta$, which is equal to the derivative of $M^{-1} X \beta$ with respect to $\rho$, we have
\begin{align}
    \frac{\partial}{\partial \rho} \ell (y | \mu, \phi, \rho) &= \sum_{i=1}^n \phi_i \sum_{j=1}^J \mu_{ij} \bigg( \ln (y_{ij}) \Big( U_{ij}  - \sum_{j'} \mu_{ij'} U_{ij'} \Big) - U_{ij} \Big( \psi(\phi_i \mu_{ij}) -   \sum_{j'} \mu_{ij'} \psi(\phi_i \mu_{ij'}) \Big) \bigg). \label{eq:derivative_ll_rho}
\end{align}
Similar to the previous subsection, to keep the presentation concise, the detailed computations of the Hessian can be found in Appendix \ref{appendix:with_spatial}. This time, the parameters are estimated with the BFGS-B algorithm \cite{byrd1995limited} due to the constraint of $\rho \in [-1,1]$.

\subsection{Multinomial distribution and cross-entropy}

The multinomial distribution is also suited to model compositional data. In this subsection, we present the multinomial regression model that will later be used as a comparison with the Dirichlet model.

Let $\tilde Y_i = (\tilde Y_{i1},\dots,\tilde Y_{iJ})$ be $n$ independent random variables  with a multinomial distribution of probabilities $p_{i1}, \dots, p_{iJ}$ with $n_i$ trials, denoted by $\tilde Y_i \sim \text{Mult}(n_i, p_{i1}, \dots, p_{iJ})$, where $\sum_j p_{ij} = 1$ and $\sum_j \tilde Y_{ij} = n_i$. Note that each $\tilde Y_{ij}$ is a non-negative integer. For an expression of the probability mass function of this distribution, refer to Appendix \ref{appendix:equivalence_crossentropy_multinomial}. 

In order to have the data summing up to 1 and belonging in the simplex $S^J$, we can define the labels $y_{ij}$ as being equal to the ratio $\tilde Y_{ij}/n_i$.  
Further, in the same spirit as the Dirichlet regression models, we can define the multinomial regression model for compositional data as in \eqref{eq:def_mu_non_spatial} and \eqref{eq:def_mu_spatial},  using
\[
p_{ij} = \begin{cases} \displaystyle
\frac{\exp([X \beta]_{ij})}{\sum_{j'=1}^J \exp([X \beta]_{ij'})},& \quad \text{for multinomial distribution},  \\
\, \vspace{-3mm}\\
\displaystyle \frac{\exp([M^{-1} X \beta]_{ij})}{\sum_{j'=1}^J \exp([M^{-1} X \beta]_{ij'})},&\quad \text{for spatial multinomial distribution.} 
\end{cases}
\]
Here, the predictions of the model are given by $\hat y_i = \hat p_i$. For the sake of simplicity, we will refer to these models as multinomial regression model (respectively, spatial multinomial regression model).
In addition, the cross-entropy between the probability vectors $y_i$ and $\hat y_i$ is given by\begin{equation}
CE_i(y_i, \hat y_i) =  - \sum_{j=1}^J y_{ij} \ln (\hat y_{ij}),
\label{eq:ce}
\end{equation}
then the cross-entropy loss over the whole dataset is the summation of all the $CE_i(y_i, \hat y_i)$ over $i$.
The regression parameters of the Dirichlet and multinomial regression models can be estimated either by maximizing the likelihood or minimizing the cross-entropy loss, potentially yielding different results.

We found that minimizing this cross-entropy function sometimes gave a better estimation of the parameters than maximization of the likelihood of the Dirichlet. This behaviour can be explained by the fact that the minimization of the cross-entropy is equivalent to the maximization of the likelihood of a multinomial distribution, given that all the $n_i$ are equal; a detailed description is provided in Appendix \ref{appendix:equivalence_crossentropy_multinomial}. Therefore, it can be inferred that if minimizing the cross-entropy yields superior results compared to maximizing the likelihood of the Dirichlet distribution, the most suitable distribution for fitting the data is likely to be multinomial rather than Dirichlet. This phenomenon may stem from the nature of the compositional data within the datasets we employed. These data may not inherently depict proportions but rather counts across various classes (e.g., number of pixels, number of votes), which are subsequently normalized by dividing them by the sum across all classes. As a consequence, the multinomial distribution is probably more suited to this kind of data.
The minimization of the cross-entropy loss function, for a model with or without spatial parameter, can be compared with the maximization of the likelihood of the Dirichlet in terms of performances (see Appendix \ref{appendix:multinomial}). Due to its simpler expression, and probably because it does not need the computation of gamma and polygamma functions, the cross-entropy is faster to optimize.
In the literature, various estimation strategies have been proposed to estimate the parameters of the spatial multinomial distribution (assuming same number of trials $n_i$ for each observation), including a maximum likelihood estimator \cite{cao2011multinomial,li2013identifying} and a Bayesian estimation strategy \cite{krisztin2022spatial}.

\subsection{Metrics}

Recall that in this section, the true probability of class $j$ for sample $i$ is represented by $y_{ij}$, and the estimated value by the model is denoted as $\hat y_{ij}$.

To evaluate and compare the performance of the spatial model and the non-spatial model, several metrics are employed. One commonly used such metric is the Akaike information criterion (AIC). By defining $k$ as the number of estimated parameters and $\hat \ell$ as the maximized log-likelihood defined in \eqref{eq:loglikelihood}, the AIC is calculated as
\begin{equation*}
    \text{AIC} = - 2 \hat \ell + 2k.
\end{equation*}
Note that smaller the values of AIC, the better the performance of the model.

Another popular metric that used in practice is the cross-entropy, given by
\begin{equation*}
    \text{Cross-entropy} = - \frac{1}{n} \sum_{i=1}^n \sum_{j=1}^J y_{ij} \log (\hat y_{ij}).
\end{equation*}
Similar to AIC, the performance of the model improves as the cross-entropy values decrease.

Additionally, the root mean squared error (RMSE) is utilized to measure model accuracy. As we are dealing with multi-class compositional vectors, the RMSE between a true vector and an estimated vector is computed as the average RMSE across all classes using
\begin{equation*}
    \text{RMSE} = \sqrt{\frac{1}{n} \frac{1}{J} \sum_{i=1}^n \sum_{j=1}^J (y_{ij} - \hat y_{ij})^2}.
\end{equation*}
Again note that smaller values of RMSE indicate better model performance.

Furthermore, the coefficient of determination, typically denoted as $R^2$, can be computed on each class. For a given class $j$,
\begin{equation}
\label{eqn:Rsquare}
    R^2_j = 1 - \frac{\sum_{i=1}^n (y_{ij} - \hat y_{ij})^2}{\sum_{i=1}^n (y_{ij} - \bar y_j)^2},
\end{equation}
where $\bar y_j$ represents the mean of the values in the $j$th class. The overall~$R^2$ value is obtained by averaging~$R^2_j$ values across all classes. Note that, although it may seem counter-intuitive, this definition allows for~$R^2$ to be negative in cases where the predicted values $\hat y_{ij}$ deviate more from the true values than the mean $\bar y_j$. It can happen in the cases when the true values are tightly clustered together (which can easily happen with compositional data), their deviation from the mean becoming very small, making the numerator in \eqref{eqn:Rsquare} close to zero. In such cases, the $R^2$ value tends to be negative or approach negative infinity. 

Finally, the metric that is probably the most suitable to compare the distance between two vectors labels is the cosine similarity, representing the cosine of the angle formed between the two vectors, given by
\begin{equation*}
\text{Cosine similarity} = \frac{1}{n} \sum_{i=1}^n \frac{y_i \hat y_i^T}{\|y_i\|_2 \, \|\hat y_i \|_2 }.
\end{equation*}

\section{Results}
\label{sec:results}

In this section, we present the results obtained from applying both the spatial lag model and the non-spatial model to various datasets. Each dataset is described in detail, along with the corresponding outcomes achieved by the respective models.

\subsection{Synthetic dataset}
\label{subsec:synthetic}

The synthetic spatially-correlated dataset is generated with 2 features and 3 classes, by varying the number of samples $n$ (50, 200, or 1000) and the values of $\rho$ (0.1, 0.5, or 0.9), which are all positive, as most often found in literature \cite{anselin1998spatial}. The values of $\beta$ and $\gamma$ are predetermined at the beginning of the simulation. In the presented results, the parameters are
\[
\beta = \stackrel{\mbox{classes}}{\begin{bmatrix} 0 & 0 & 0.1 \\ 0 & 1 & -2 \\ 0 & -1 & -2
\end{bmatrix}}\mbox{features }, \hspace{1cm} \gamma = \begin{bmatrix} 2 \\ 3
\end{bmatrix}.
\]
These specific values were selected to achieve certain desirable properties in the generated synthetic data.
In particular, the value of $\beta$ is chosen to ensure that the classes are balanced in $\mu$, i.e., that no class is significantly more frequent or rarer than others. With such a value for $\beta$, the data are generated with class distributions that are evenly spread, avoiding any class dominance or extreme imbalances.
Regarding $\gamma$, its chosen value is intended to ensure that the precision parameter $\phi$, which controls the spread or dispersion of the class probabilities, is sufficiently high, so that the distribution of the points is relatively concentrated around their class probabilities. This way, the class patterns are more distinguishable and well-defined.

Initially, $n$ samples are randomly drawn from a multivariate normal distribution with two covariates, producing the features matrix $X \in \mathbb R^{n \times 2}$. The matrix $Z$ is constructed with one covariate drawn from a uniform distribution. The matrix $W$ is built such that $W_{ij} = 1/k$ if $1 \leq |i-j| \leq k$, where $k$ is the number of neighbors, otherwise $W_{ij} = 0$.
In our simulation, we specifically considered $k=5$ neighbors, but additional simulations involving more neighbors suggested comparable results. 

An alternative approach, based on \cite{beron2004probit, calabrese2014estimators}, is to construct the matrix $W$ by selecting a radius of a specific size (depending on the value of $\rho$) around each point in $X$, and to consider all the values falling within this radius as neighbors of the corresponding point. This method, however, did not seem to recreate the behaviour of a spatial effect, as it only repeated the information already present in the matrix $X$.

Then, from the matrices $X$ and $Z$ and the parameters $\beta$ and $\gamma$, we compute $\mu$ and $\phi$ that are then used to produce $\alpha$. The response matrix $Y$ is finally generated by drawing in the Dirichlet distribution of parameter $\alpha_i$ for each row $i$.

We repeat 100 times the following experiment: the data are created and the bias of the estimated parameters is computed. The results for each of the three $\rho$ are presented in Tables \ref{table:scores_synthetic_01}, \ref{table:scores_synthetic_05} and \ref{table:scores_synthetic_09}.

\begin{table}
    \small 
	\begin{center}
    \caption{Estimated bias (averaged over 
    $n=100$ replications) of the estimated parameters of the Dirichlet model in the context of Dirichlet generated data with $\rho=0.1$. Standard deviations are presented within parenthesis, and mean squared errors within square brackets.}
    \label{table:scores_synthetic_01}
		\begin{tabular}{|c|c|c|c|c|c|c|}
			\hline
   & \multicolumn{3}{c|}{Spatial} & \multicolumn{3}{c|}{Not spatial} \\
			\hline
			Parameter & $n=50$ & $n=200$ & $n=1000$ & $n=50$ & $n=200$ & $n=1000$ \\
			\hline
			\multirow{2}{*}{$\beta_{01}$} & 0.013 (0.059) & 0.012 (0.032) & 0.010 (0.016) & 0.014 (0.067) & 0.012 (0.036) & 0.011 (0.018) \\
			 & [0.004] & [0.001] & [0.0] &  [0.005] &  [0.001] &  [0.0] \\
			\hline
			\multirow{2}{*}{$\beta_{02}$} & 0.038 (0.087) & 0.035 (0.034) & 0.031 (0.016) & 0.051 (0.106) & 0.055 (0.046) & 0.047 (0.021) \\
			 & [0.009] & [0.002] & [0.001] &  [0.014] &  [0.005] &  [0.003] \\
			\hline
			\multirow{2}{*}{$\beta_{11}$} & -0.062 (0.079) & -0.037 (0.038) & -0.019 (0.017) & -0.067 (0.081) & -0.040 (0.038) & -0.021 (0.017) \\
			 & [0.010] & [0.003] & [0.001] &  [0.011] &  [0.003] &  [0.001] \\
			\hline
			\multirow{2}{*}{$\beta_{12}$} & 0.277 (0.12) & 0.185 (0.051) & 0.141 (0.023) & 0.288 (0.127) & 0.198 (0.055) & 0.158 (0.024) \\
			 & [0.091] & [0.037] & [0.020] &  [0.099] &  [0.042] &  [0.026] \\
			\hline
			\multirow{2}{*}{$\beta_{21}$} & 0.042 (0.083) & 0.015 (0.035) & 0.004 (0.014) & 0.045 (0.081) & 0.017 (0.035) & 0.007 (0.015) \\
			 & [0.009] & [0.001] & [0.0] &  [0.009] &  [0.002] &  [0.0] \\
			\hline
			\multirow{2}{*}{$\beta_{22}$} & 0.197 (0.108) & 0.123 (0.05) & 0.088 (0.023) & 0.203 (0.105) & 0.131 (0.053) & 0.101 (0.025) \\
			 & [0.051] & [0.018] & [0.008] &  [0.052] &  [0.02] &  [0.011] \\
			\hline
			\multirow{2}{*}{$\gamma_0$} & 0.534 (0.312) & 0.353 (0.139) & 0.269 (0.066) & 0.522 (0.32) & 0.349 (0.144) & 0.264 (0.066) \\
			 & [0.382] & [0.144] & [0.077] &  [0.375] &  [0.143] &  [0.074] \\
			\hline
			\multirow{2}{*}{$\gamma_1$} & -0.354 (0.598) & -0.324 (0.246) & -0.294 (0.105) & -0.453 (0.612) & -0.448 (0.266) & -0.416 (0.11) \\
			 & [0.483] & [0.165] & [0.098] &  [0.58] &  [0.272] &  [0.185] \\
			\hline
			\multirow{2}{*}{$\rho$} & -0.01 (0.052) & -0.003 (0.023) & -0.0 (0.009) & / & / & / \\
			 & [0.003] & [0.001] & [0.0] & / & / & / \\
			\hline
		\end{tabular}
	\end{center}
\end{table}

\begin{table}
    \small
	\begin{center}
    \caption{Estimated bias (averaged over 
    $n=100$ replications) of the estimated parameters of the Dirichlet model in the context of Dirichlet generated data with $\rho=0.5$. Standard deviations are presented within parenthesis, and mean squared errors within square brackets.}
    \label{table:scores_synthetic_05}
		\begin{tabular}{|c|c|c|c|c|c|c|}
			\hline
        &  \multicolumn{3}{c|}{Spatial} & \multicolumn{3}{c|}{Not spatial} \\ \hline
			Parameter & $n=50$ & $n=200$ & $n=1000$ & $n=50$ & $n=200$ & $n=1000$ \\ \hline
			\multirow{2}{*}{$\beta_{01}$} & 0.009 (0.042) & 0.011 (0.02) & 0.007 (0.009) & 0.029 (0.157) & 0.03 (0.079) & 0.028 (0.037) \\
			 & [0.002] & [0.001] & [0.0] &  [0.026] &  [0.007] &  [0.002] \\
			\hline
			\multirow{2}{*}{$\beta_{02}$} & 0.012 (0.047) & 0.023 (0.017) & 0.017 (0.009) & 0.116 (0.312) & 0.184 (0.147) & 0.144 (0.066) \\
			 & [0.002] & [0.001] & [0.0] &  [0.111] &  [0.055] &  [0.025] \\
			\hline
			\multirow{2}{*}{$\beta_{11}$} & -0.06 (0.083) & -0.037 (0.037) & -0.024 (0.016) & -0.163 (0.122) & -0.17 (0.057) & -0.155 (0.027) \\
			 & [0.011] & [0.003] & [0.001] &  [0.042] &  [0.032] &  [0.025] \\
			\hline
			\multirow{2}{*}{$\beta_{12}$} & 0.356 (0.123) & 0.221 (0.061) & 0.176 (0.029) & 0.682 (0.205) & 0.622 (0.099) & 0.598 (0.049) \\
			 & [0.142] & [0.052] & [0.032] &  [0.507] &  [0.397] &  [0.36] \\
			\hline
			\multirow{2}{*}{$\beta_{21}$} & 0.051 (0.067) & 0.016 (0.041) & 0.01 (0.015) & 0.142 (0.123) & 0.129 (0.055) & 0.129 (0.028) \\
			 & [0.007] & [0.002] & [0.0] &  [0.035] &  [0.02] &  [0.017] \\
			\hline
			\multirow{2}{*}{$\beta_{22}$} & 0.246 (0.108) & 0.145 (0.049) & 0.113 (0.022) & 0.511 (0.202) & 0.472 (0.088) & 0.462 (0.042) \\
			 & [0.072] & [0.023] & [0.013] &  [0.302] &  [0.23] &  [0.216] \\
			\hline
			\multirow{2}{*}{$\gamma_0$} & 0.523 (0.314) & 0.418 (0.153) & 0.285 (0.062) & 0.015 (0.364) & -0.178 (0.164) & -0.337 (0.087) \\
			 & [0.372] & [0.198] & [0.085] &  [0.132] &  [0.059] &  [0.121] \\
			\hline
			\multirow{2}{*}{$\gamma_1$} & -0.363 (0.637) & -0.395 (0.289) & -0.313 (0.109) & -1.828 (0.749) & -2.0 (0.293) & -1.92 (0.137) \\
			 & [0.538] & [0.24] & [0.11] &  [3.903] &  [4.086] &  [3.704] \\
			\hline
			\multirow{2}{*}{$\rho$} & -0.005 (0.032) & -0.002 (0.011) & 0.0 (0.005) & / & / & / \\
			 & [0.001] & [0.0] & [0.0] & / & / & / \\
			\hline
		\end{tabular}
	\end{center}
\end{table}

\begin{table}
    \small
	\begin{center}
    \caption{Estimated bias (averaged over 
    $n=100$ replications) of the estimated parameters of the Dirichlet model in the context of Dirichlet generated data with $\rho=0.9$. Standard deviations are presented within parenthesis, and mean squared errors within square brackets.}
    \label{table:scores_synthetic_09}
		\begin{tabular}{|c|c|c|c|c|c|c|}
			\hline
			  &  \multicolumn{3}{c|}{Spatial} & \multicolumn{3}{c|}{Not spatial} \\ \hline
			Parameter & $n=50$ & $n=200$ & $n=1000$ & $n=50$ & $n=200$ & $n=1000$ \\
			\hline
			\multirow{2}{*}{$\beta_{01}$} & 0.007 (0.076) & 0.011 (0.021) & 0.01 (0.011) & 0.096 (0.596) & 0.019 (0.24) & 0.044 (0.097) \\
			 & [0.006] & [0.001] & [0.0] &  [0.364] &  [0.058] &  [0.011] \\
			\hline
			\multirow{2}{*}{$\beta_{02}$} & -0.022 (0.126) & -0.019 (0.03) & -0.024 (0.017) & 0.303 (1.096) & 0.298 (0.381) & 0.206 (0.165) \\
			 & [0.016] & [0.001] & [0.001] &  [1.292] &  [0.234] &  [0.069] \\
			\hline
			\multirow{2}{*}{$\beta_{11}$} & -0.418 (0.335) & -0.423 (0.194) & -0.39 (0.112) & -0.653 (0.254) & -0.696 (0.113) & -0.673 (0.05) \\
			 & [0.287] & [0.217] & [0.165] &  [0.491] &  [0.496] &  [0.455] \\
			\hline
			\multirow{2}{*}{$\beta_{12}$} & 1.193 (0.481) & 1.185 (0.345) & 1.148 (0.218) & 1.588 (0.27) & 1.559 (0.088) & 1.531 (0.049) \\
			 & [1.653] & [1.523] & [1.365] &  [2.596] &  [2.437] &  [2.345] \\
			\hline
			\multirow{2}{*}{$\beta_{21}$} & 0.356 (0.271) & 0.368 (0.196) & 0.359 (0.114) & 0.611 (0.233) & 0.655 (0.108) & 0.652 (0.052) \\
			 & [0.2] & [0.174] & [0.142] &  [0.427] &  [0.441] &  [0.428] \\
			\hline
			\multirow{2}{*}{$\beta_{22}$} & 0.997 (0.434) & 1.003 (0.347) & 0.976 (0.218) & 1.444 (0.253) & 1.446 (0.114) & 1.444 (0.056) \\
			 & [1.182] & [1.127] & [1.0] &  [2.149] &  [2.104] &  [2.089] \\
			\hline
			\multirow{2}{*}{$\gamma_0$} & 0.411 (0.679) & -0.257 (0.564) & -0.414 (0.419) & -1.24 (0.645) & -1.924 (0.149) & -2.174 (0.072) \\
			 & [0.629] & [0.384] & [0.347] &  [1.952] &  [3.724] &  [4.731] \\
			\hline
			\multirow{2}{*}{$\gamma_1$} & -1.909 (0.759) & -2.151 (0.547) & -2.167 (0.323) & -2.717 (0.474) & -2.787 (0.128) & -2.779 (0.065) \\
			 & [4.218] & [4.927] & [4.8] &  [7.606] &  [7.784] &  [7.729] \\
			\hline
			\multirow{2}{*}{$\rho$} & -0.011 (0.051) & -0.006 (0.031) & -0.004 (0.017) & / & / & / \\
			 & [0.003] & [0.001] & [0.0] & / & / & / \\
			\hline
		\end{tabular}
	\end{center}
\end{table}

The results show that, in regard to the non-spatial parameters $\beta$ and $\gamma$,  both models demonstrate similar behavior, with their bias, variance, and mean squared error being asymptotically unbiased. In the spatial lag models, we also observe the expected behavior, where the bias and mean squared error of the estimated $\hat \rho$ decrease as the number of samples increases. It is worth noting that when $\hat \rho$ is biased (which occurs when the sample size $n$ is small), the bias is negative, suggesting that the model tends to underestimate the spatial correlation strength. 

The prediction accuracy of the models is assessed as follow.
For each value of $\rho$, we create the test set by generating 1000 new data points using the true parameters $\beta^*$. On this test data, the true value $\mu^*$ is computed. Then, we compute $\hat\mu$ using the parameters $\hat \beta$ estimated from the $n=1000$ simulation. To evaluate the difference between $\mu^*$ and each $\hat\mu$, metrics such as $R^2$, RMSE, cross-entropy and cosine similarity are utilized. The results are presented in Table \ref{table:scores_test_synthetic}.
For a low spatial correlation ($\rho=0.1$), both models perform equally well. However, as the spatial correlation increases ($\rho=0.5$ and $\rho=0.9$), the performances of the non-spatial model decrease, and they are outperformed by the spatial lag model across all metrics. 
Interestingly, the spatial lag model has best performance when the spatial correlation strength is moderately high ($\rho=0.5$), exhibiting slightly better results compared to the scenario with an extremely high correlation strength ($\rho=0.9$). This suggests that the spatial lag model is most effective when there is a moderate level of spatial dependence, and its performance may plateau or decline at extremely high spatial correlation levels.

\begin{table}[h!]
  \centering
  \caption{Performance measures on the test set comparing $\mu^*$ and the estimated $\hat \mu$ (computed with the parameters estimated with $n=1000$). The results are displayed as mean on the 100 iterations (standard deviation).}
  \label{table:scores_test_synthetic}
  \begin{tabular}{|c|l|c|c|c|c|}
    \hline
    & \textbf{Model} & $\boldsymbol{R^2}$ & \textbf{RMSE} & \textbf{Cross-entropy} & \textbf{Cos similarity} \\
    \hline
    \multirow{2}{*}{$\rho=0.1$} & Not spatial & 0.9309 ($<10^{-4}$) & 0.0739 ($<10^{-4}$) & 0.6660 (0.0001) & 0.9855 ($<10^{-4}$) \\
    & Spatial & 0.9335 ($<10^{-4}$) & 0.0723 ($<10^{-4}$)  & 0.6648 (0.0001) & 0.9862 ($<10^{-4}$) \\ \hline
    \multirow{2}{*}{$\rho=0.5$} & Not spatial & 0.8311 (0.0001) & 0.1249 ($<10^{-4}$) & 0.6845 (0.0001) & 0.9610 ($<10^{-4}$)\\
    & Spatial & 0.9408 ($<10^{-4}$) & 0.0705 ($<10^{-4}$) & 0.6275 (0.0002) & 0.9872 ($<10^{-4}$) \\ \hline
    \multirow{2}{*}{$\rho=0.9$} & Not spatial & 0.2073 (0.0025) & 0.3257 (0.0001) & 0.9033 (0.0005) & 0.7764 (0.0001)  \\
    & Spatial & 0.9011 (0.0026) & 0.1097 (0.0008) & 0.4414 (0.0026) & 0.9776 (0.0001) \\
    \hline
  \end{tabular}
\end{table}

We then also try to retrieve the parameters from a multinomial model. Note that when the number of trials is equal for each observation, the estimation of the multinomial model using the maximum likelihood is equivalent to minimizing the cross-entropy. The results, presented in Appendix \ref{appendix:multinomial}, were similar to those of the Dirichlet model. However, we observed that when the $\gamma$ parameter was decreased, leading to less distinguishable classes, the performances of the Dirichlet model significantly deteriorated compared to the multinomial model.

Additionally, we conducted experiments where the labels were generated using a multinomial distribution instead of a Dirichlet distribution. In these cases, the parameter $n_i$, representing the number of trials for each observation $i$, was drawn from a discrete uniform distribution between 100 and 10000. In Appendix \ref{appendix:synthetic_multinomial}, more details are given on how the data are generated with a multinomial distribution, and the performances of both the multinomial and the Dirichlet models are compared. As expected, in that case, the multinomial model performs better than the Dirichlet model, especially under a high spatial correlation.

\subsection{Arctic Lake}
\label{subsec:arctic}

The Arctic Lake dataset \cite{coakley1968sedimentation} provides compositional data in terms of sand, silt, and clay percentages for 39 sediment samples taken at various water depths in an Arctic lake. The goal is to quantify the extent to which water depth influences the compositional patterns of the sediment samples. We propose here two Dirichlet regression models: an order 1 model with a single predictor variable (the depth) along with an intercept term, and an order 2 model that includes an intercept term, the depth variable, and its squared value as additional predictor. The models are estimated using the maximum likelihood approach.

Due to the limited size of the dataset, we employ a Leave One Out Cross Validation (LOOCV) strategy. We iteratively exclude one sample (the $k$-th sample) from the dataset, and use the remaining samples to estimate the parameters of the models. We then compute the predicted odd values $\hat\mu_k$ for the excluded sample, and evaluate its proximity to the true compositional label using metrics such as $R^2$, RMSE, cross-entropy, AIC and cosine similarity. The mean and standard deviation of these metrics are reported in Table~\ref{table:scores_arctic_lake}.

\begin{table}[h]
\begin{adjustwidth}{-1cm}{0cm}
\centering
\caption{Performance measures for the models on Arctic Lake dataset computed with a Dirichlet regression model using a LOOCV strategy. The presented results are the mean value, and standard deviation within parenthesis.}
\label{table:scores_arctic_lake}
\begin{tabular}{|l|l|c|c|c|c|c|c|}
\hline
\multicolumn{2}{|c|}{\textbf{Model}} & $\boldsymbol{R^2}$ & \textbf{RMSE} & \textbf{Cross-entropy} & \textbf{AIC} & \textbf{Cos similarity} \\
\hline
& without spatial & 0.5887 (0.015) & 0.1015 (0.0018) & 0.9134 (0.0027) & -141.6 (2.3) & 0.9665 (0.0012) \\
Order 1 & spatial contiguity & \textbf{0.6323} (0.0147) & \textbf{0.0951} (0.0017) & \textbf{0.9064} (0.0029) & \textbf{-149.5} (2.3) & \textbf{0.9706} (0.0011)\\
& spatial distance & 0.5893 (0.0148) & 0.1014 (0.0018) & 0.9134 (0.0027) & -139.7 (2.2) & 0.9665 (0.0012) \\
\hline
& without spatial& 0.6784 (0.0144) & 0.0881 (0.0019) & 0.8993 (0.0032) & -168.1 (3.3) & 0.9743 (0.0011) \\
Order 2 & spatial contiguity & \textbf{0.6943} (0.0144) & \textbf{0.0858} (0.0019) & \textbf{0.8974} (0.0033) & \textbf{-170.6} (3.6) & \textbf{0.9755} (0.0011) \\
& spatial distance & 0.6863 (0.015) & 0.0871 (0.002) & 0.8982 (0.0032) & -167.1 (3.5) & 0.9747 (0.0012) \\
\hline
\end{tabular}
\end{adjustwidth}
\end{table}

The results suggest that the utilization of spatial information leads to slight improvements in model performance. However, in terms of variability, the difference may not be statistically significant. This lack of significance could be attributed to the spatial information being derived solely from the depth variable, resulting in the absence of any new information being introduced. Instead, the data is essentially replicated in a different manner.

\subsection{Maupiti}
\label{subsec:maupiti}

Maupiti Island, situated in the Society archipelago of French Polynesia, is an island spanning approximately 8km by 8km in size. The dataset originates from an expert-driven mapping process of Maupiti Island, based on various field observation campaigns \cite{sous2020small}. Details on how the dataset has been created can be found on a previous study on compositional data \cite{nguyen2023smote}. Each of the 2091 samples of the dataset are segments created via Felzenszwalb's segmentation \cite{felzenszwalb2004efficient} on the RGB Pléiades satellite image (Pléiades © CNES 2021, Distribution AIRBUS DS), each of them having 16 features (4 statistical moments on 4 different bands) and a compositional data label composed of 4 classes: coral, sand, shorereef, and mixed. The label is derived from the expert's ground-truth map: after the segmentation process, we count the pixels assigned to each class within each segment, and divide it by the total number of pixels within the respective segment. Note that in this case the number of pixels in each segment is different.

The neighbors matrix $W$ is created as a distance-based matrix between each segment and its neighbor. Note that we also experimented with a contiguity-based matrix, but it yielded inferior performances. Since our analysis specifically focuses on the shallow regions of the lagoon, which offer clearer and more easily interpretable imagery, some segments are excluded from the analysis, which in turn can result in certain segments not having any neighbors.

The matrix $Z$ is chosen as being a copy of the features matrix $X$, as this choice yielded the best results compared to the case where $Z$ is only an intercept.

We use the Dirichlet models to retrieve the parameters and then, an estimated $\hat\mu$ is computed and compared with the real $\mu$. The metrics to compare them are reported in Table \ref{table:scores_maupiti}. The spatial correlation parameter $\hat \rho$ is estimated to be 0.93 with this dataset, indicating a high spatial correlation between these segments.

\begin{table}[h]
\centering
\caption{Performance measures for the Dirichlet models on Maupiti dataset.}
\label{table:scores_maupiti}
\begin{tabular}{|l|c|c|c|c|c|}
\hline
\textbf{Model} & $\boldsymbol{R^2}$ & \textbf{RMSE} & \textbf{Cross-entropy} &            \textbf{AIC} & \textbf{Cos similarity} \\
\hline
Without spatial & 0.265 & 0.297 & 0.815 & -73873 & 0.786 \\ \hline
With spatial & 0.441 & 0.261 & 0.675 & -74170 & 0.848 \\ \hline
\end{tabular}
\end{table}

The spatial lag model consistently outperforms the non-spatial model across all metrics. However, it is worth noting that the performance of the spatial lag model remains relatively low, with an $R^2$ value of 0.441 and a cosine similarity of 0.848. This suggests that the use of the Dirichlet distribution may not be well-suited for this specific dataset. It is currently unclear whether this limitation is specific to the unique characteristics of the Maupiti dataset or if the Dirichlet distribution generally lacks suitability for datasets involving geographical maps. Furthermore, Table \ref{table:scores_maupiti_crossentropy} reveals that the multinomial regression models estimated using the cross-entropy loss on this dataset yields significantly better performances than the Dirichlet regressor, further confirming that the Dirichlet distribution may not be well suited for this dataset.

\begin{table}[h]
\centering
\caption{Performance measures for the mutinomial regression models on Maupiti dataset using the cross-entropy loss}
\label{table:scores_maupiti_crossentropy}
\begin{tabular}{|l|c|c|c|c|c|}
\hline
\textbf{Model} & $\boldsymbol{R^2}$ & \textbf{RMSE} & \textbf{Cross-entropy} &            \textbf{Cos similarity} \\
\hline
Without spatial & 0.636 & 0.221 & 0.455 & 0.868 \\ \hline
With spatial & 0.820 & 0.160 & 0.307 & 0.925 \\ \hline
\end{tabular}
\end{table}

\begin{figure}[h]
    \centering
    \includegraphics[width=15cm]{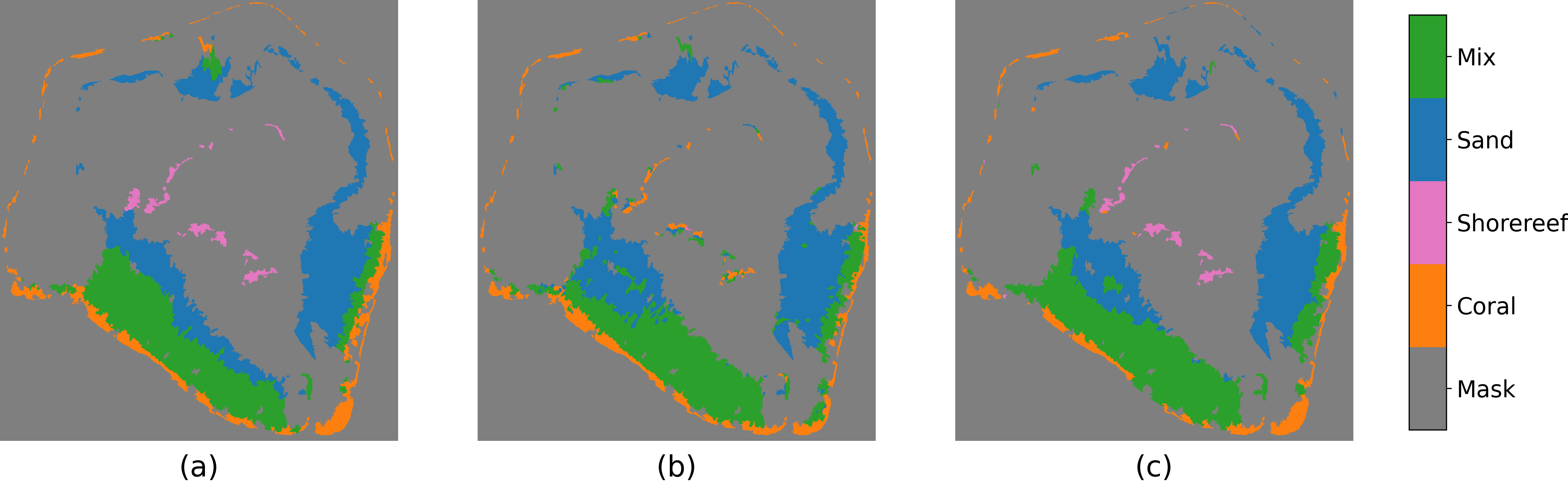}
    \caption{Maps created by using (a) the real labels, (b) the labels computed with the non-spatial Dirichlet model, and (c) the labels computed with the spatial Dirichlet model. The pixel-wise accuracy between maps (a) and respectively (b) and (c) is 0.752 and 0.832.}
    \label{fig:maps_maupiti}
\end{figure}

Furthermore, we compute the maximum a posteriori (MAP) by taking the argmax of the compositional data label for each segment and assign it as the class of the segment. This process allows for the creation of a visually interpretable map. The ground-truth map resulting from the expert-based mapping is depicted in Figure \ref{fig:maps_maupiti}a, along with the maps obtained from the predictions of the non-spatial (Figure \ref{fig:maps_maupiti}b) and spatial Dirichlet models (Figure \ref{fig:maps_maupiti}c). To evaluate the accuracy of these models, pixel-wise comparisons are conducted to count the number of correctly assigned pixels, which is then divided by the total number of classified pixels, excluding masked pixels. The accuracy achieved by the non-spatial model is 0.752, while the spatial lag model achieves an accuracy of 0.832, further confirming that the spatial lag model provides superior results.

\subsection{Elections}
\label{subsec:elections}

This dataset present the votes at the French departmental election of 2015 in the Occitanie region \cite{goulard2017predictions, nguyen2022analyzing}, for $n = 207$ cantons. For each canton, the voting distribution (initially between 15 political parties) is categorized into three major political movements: left, right, and extreme right. In our study, we utilized 25 distinct social indicators as features, including age categories, employment fields, and education level, among others. Initially, the dataset consisted of 283 cantons, but any cantons where one of the classes was not present were removed. This resulted in the exclusion of 76 points, which accounts for 27\% of the data.

The spatial weights matrix $W$ is computed based on the geographic proximity of each canton's center. Specifically, two cases are considered. In the first case, the contiguity-based, we consider the 5 nearest neighboring cantons, determined by their center-to-center distances. In the second case, the distance-based, the inverse of the distance between each canton and the others is considered, with a cut-off at a certain value that minimizes the average number of neighbors and to ensure that each canton has at least one neighbor. This cut-off gives 12 neighbors on average.

The matrix $Z$ is chosen as being a sole intercept, as this choice yielded the best results compared to the case where $Z$ is a copy of the features matrix $X$.

Then, for the three models (non-spatial and the two spatials), we use the maximum likelihood estimator to retrieve the parameters and compute the performance with our usual metrics. Results are reported in Table \ref{table:scores_elections}. The estimated spatial correlation coefficient $\hat \rho$ is 0.97 (resp. 0.91) with the distance-based (resp. contiguity-based) matrix.

\begin{table}[ht]
\centering
\caption{Performance measures for the Dirichlet models on Elections dataset.}
\label{table:scores_elections}
\begin{tabular}{|l|c|c|c|c|c|c|}
\hline
\textbf{Model} & $\boldsymbol{R^2}$ & \textbf{RMSE} & \textbf{Cross-entropy} & \textbf{AIC} &  \textbf{Cos similarity} \\
\hline
No spatial & 0.487 & 0.080 & 1.048 & -862.1 & 0.975 \\ \hline
Spatial (contiguity) & 0.582 & 0.072 & 1.042 & -947.4 & 0.979 \\ \hline
Spatial (distance) & \textbf{0.602} & \textbf{0.070} & \textbf{1.041} & \textbf{-965.1} & \textbf{0.980} \\ \hline
\end{tabular}
\end{table}

The spatial lag models perform better than the non-spatial model across all evaluation metrics, excepted for the AIC which is slightly better for the non-spatial model. Besides, the distance-based spatial lag model performs slightly better than the contiguity-based. Additionally, we attempted to make predictions using our model through a 10-fold cross-validation technique. In this approach, 90\% of the data were used to estimate the model parameters, while the remaining 10\% (corresponding to 21 values) were reserved for testing the model's performance. However, we observed extremely poor performance on the test set, indicating that the spatial lag model is highly sensitive to missing values. This could be attributed to the fact that 27\% of the initial data was already missing, and further data removal might have rendered the spatial information irrelevant. Notably, in a previous study \cite{goulard2017predictions}, the authors were able to successfully make predictions using the entire dataset of 283 cantons.

An important observation with this dataset is that the multinomial distribution and the minimization of the cross-entropy yielded similar results compared to the Dirichlet distribution.

\section{Discussion}
\label{sec:discussion}

It is important to remember that the minimization of cross-entropy is equivalent to maximizing the likelihood of a multinomial distribution when the number of trial in each observation the same. However, even when they are different, the results remain quite similar. We conducted several tests by generating synthetic datasets with varying sample sizes, and interestingly, the performances of a multinomial regressor using sample size information were comparable to those of a regressor assuming similar sample sizes. Thus, we will refer to both cases as the ``multinomial model'', which encompasses the minimization of cross-entropy. Additionally, when we evaluated a Dirichlet regression model on these multinomial generated data, we found that it performed similarly to the multinomial model, excepted in the cases of extremely high spatial correlation, where the multinomial approach was slightly better. 

For the Dirichlet regression models, the analysis of the synthetic dataset reveals that both the spatial and non-spatial estimators appear to be asymptotically unbiased. As shown in Table \ref{table:scores_test_synthetic}, when the spatial correlation in the data is low ($\rho=0.1$), there is no significant difference in the performances between the spatial and non-spatial models. This finding was further confirmed when we generated data without spatial dependencies, and the spatial lag model accurately retrieved the parameters, estimating that $\hat \rho$ was not significantly different from 0.
On the other hand, when the spatial correlation is high ($\rho=0.5$ or 0.9), the spatial lag model outperforms the non-spatial one. Notably, the spatial lag model exhibits better performance under moderately high spatial correlation ($\rho=0.5$) compared to extremely high spatial correlation ($\rho=0.9$). This behavior may be attributed to the challenge of distinguishing between spatial patterns and the true underlying relationships between the variables when the spatial correlation is exceedingly strong. However, interestingly, under the extremely high spatial correlation scenario, the multinomial model performs slightly better than the Dirichlet model for the prediction task, as observed when comparing Tables \ref{table:scores_test_synthetic} and \ref{table:scores_test_synthetic_ce}. This suggests that the multinomial regression model may be more suited to handle high spatial correlation scenarios.

Overall, the analysis of the synthetic dataset reveals that the performance of the spatial Dirichlet model is influenced by the level of spatial correlation present in the data. While the spatial lag model demonstrates improved estimation in cases of spatially correlated data, it still exhibits non-optimal performance when dealing with strong spatial correlations. This behavior could partially explain the suboptimal results of the Dirichlet model on the Maupiti dataset (Table \ref{table:scores_maupiti}), where the estimated $\hat \rho$ was 0.93. However, as the model performs well on the Elections dataset, with a high estimated $\hat \rho$ ($>0.9$), other factors may contribute to the poor performances on Maupiti dataset.
Furthermore, the fact that the minimization of the cross-entropy yielded higher performances (Table \ref{table:scores_maupiti_crossentropy}) raises questions about whether the Dirichlet distribution was truly suitable for this particular dataset. 
Additionally, our experiments with the multinomial distribution taking into account the sample sizes, did not outperform the cross-entropy approach when assuming equal sample size. This observation aligns with the findings from the synthetic data analysis mentioned earlier in this section. 

From our analysis of the real datasets, several important conclusions can be drawn regarding the impact of spatial information on model performance. We observed that incorporating spatial information can significantly improve results when the spatial information truly represents additional data, rather than being derived from existing data. For instance, in the case of the Arctic Lake dataset, when we attempted to recreate the spatial weight matrix $W$ using only the available covariate (depth), there was no significant improvement in model performance (Section \ref{subsec:arctic}). This emphasizes the importance of incorporating genuinely new spatial information to achieve better results.

Furthermore, when a Dirichlet distribution is well-suited to the dataset, as demonstrated in the Elections dataset (Section \ref{subsec:elections}), our SAR Dirichlet model outperforms the non-spatial model. Notably, the spatial lag model exhibits robust performance even in the presence of missing data, with 27\% of the initial data missing. However, when a large amount of data is missing, as observed during a 10-fold cross-validation, the model's performance tends to degrade.

In our analysis of real-life datasets, we observed that the spatial weights matrix $W$ performed better when defined as distance-based rather than contiguity-based. However, it is essential to acknowledge that this result may not be generalized to every scenario, and it might be specific to the datasets we examined. We have not been able to find comprehensive studies in literature providing a definitive analysis to determine the best type of spatial weights matrix for all cases.

In the presented work, we estimated the parameters of the models through a probabilistic approach. This approach offers distinct advantages, primarily by providing quantifiable measures of uncertainty, including p-values, confidence intervals, and enabling statistical inference. 
To evaluate the significance of the spatial parameter $\rho$, we conducted Wald tests and Log-ratio tests (LRT). However, our analysis revealed that in the context of our SAR Dirichket model, these tests did not effectively control the significance level $\alpha$.
Furthermore, we extended our investigation to encompass general linear SAR models and encountered similar issues with controlling the significance level for the spatial parameter $\rho$. Notably, the tests applied did not maintain the desired $\alpha$ level. It's worth noting that our exploration into the literature did not yield studies that have extensively examined this phenomenon, leaving us uncertain about whether the observed effects are due to a potential oversight in our methodology or if they reflect broader trends in SAR models. Further research and investigation are necessary to shed light on this matter.

An additional enhancement for our model involves optimizing computational efficiency, a crucial aspect given the matrix inversion requirement. Specifically, we executed the models on an Intel i5 processor with 8GB of RAM. The computations for the Maupiti dataset (4 classes, $n=2091$) completed within 1 second for the non-spatial model, but extended to 38 minutes for the spatial lag model. On the Elections dataset, which is smaller than Maupiti (3 classes, $n=207$), the non-spatial model run in less than 1 second, while the spatial lag model took more than 6 minutes. The whole simulation for synthetic data (3 classes, 100 iterations with 3 different sample 
 sizes $n=50$, $n=200$, $n=1000$) required 37 minutes for $\rho=0.1$, 46 minutes for $\rho=0.5$ and 1 hour and 10 minutes for $\rho=0.9$, proving that the running time depends on the spatial correlation strength. Moreover, the most important factor is the number of samples $n$: for instance, for $\rho=0.9$, one iteration takes 300 milliseconds with $n=50$, two seconds with $n=200$ and 20 seconds with $n=1000$. These computational times prompt exploration into alternative techniques and approximations for the matrix inverse.

\section{Conclusion}
\label{sec:conclusion}

Our study demonstrates that incorporating spatial dependencies in a Dirichlet model leads to improved performance when dealing with datasets featuring compositional labels. Our findings from the real-life datasets reveal that a distance-based spatial weight matrix tends to yield better results compared to a contiguity-based matrix. These results underscore the potential advantages of spatial modeling, especially in scenarios where the Dirichlet distribution is well-suited to the data.

The results obtained from the synthetic dataset provide some insights into the behavior of the SAR Dirichlet model.  While the spatial lag model outperforms the non-spatial model in spatially correlated data, the spatial lag model does not perform optimally under extremely high spatial correlation and provides better results when the spatial correlation is moderate ($\rho = 0.5$).

Overall, our study highlights the importance of considering spatial information when it provides meaningful additional context, as it can significantly enhance the model's effectiveness. It also emphasizes the potential impact of missing data, which should be carefully addressed to avoid adverse effects on model performance.

Moreover, our findings suggest that, in general, the multinomial model appears to be better suited for handling compositional data compared to the Dirichlet model, especially in the cases where the class patterns are not highly distinguishable. Throughout our analysis, we did not encounter instances where the Dirichlet model significantly outperformed the multinomial model. However, we believe that this behaviour may be caused by the fact that the compositional data present in our datasets are not naturally a probability, but a ratio (the count of the number of pixels, or the number of votes). The multinomial distribution might be naturaly more suited to handle this case. Further investigations are required to determine if it holds true across a broader range of scenarios and datasets.

\section*{Acknowledgements}

Funding was provided by the Energy Environment Solutions (E2S-UPPA) consortium
through an international research chair. The authors would like to thank GLADYS  (\url{https://www.gladys-littoral.org/}) and Damien Sous and Samuel Meulé for providing the Maupiti data, and the research team at Toulouse School of Economics for providing the elections data.

\printbibliography

\pagebreak

\appendix

\section{Computation on Dirichlet Distribution without Spatial Lag} 
\label{appendix:without_spatial}

Let $K$ be the number of features, $J$ be the number of classes and $n$ be the number of samples. We have the features matrix $X \in \mathbb R^{n \times K}$, the label matrix $Y \in \mathbb R^{n \times J}$ and the matrix of parameters $\alpha \in \mathbb R^{n \times J}$.
To make the computations easier, we split the loglikelihood into three parts $A_i$, $B_i$ and $C_i$ as follows.

\begin{equation}
\label{eqn:loglikelihood-again}
    \ell(y | \mu, \phi) = \sum_{i=1}^n \left(
    \underbrace{\vphantom{\sum_j} \ln \Gamma ( \phi_i )}_{A_i}
    - \underbrace{\sum_j \ln ( \Gamma(\phi_i \mu_{ij}) )}_{B_i}
    + \underbrace{\sum_j ((\phi_i \mu_{ij}-1) \ln (y_{ij})) }_{C_i}
    \right).
\end{equation}
Note that $\mu$ is a function of $\beta$ and $\phi$ is a function of $\gamma$. Below, in Sections~\ref{sec:first-order-without} and \ref{sec:second-order-without} we  respectively provide computations of the first order and the second order derivatives of the loglikelihood function with respect to $\beta$ and $\gamma$.

\subsection{First order derivative}
\label{sec:first-order-without}

To compute the gradient of the loglikelihood, we compute the derivatives of $A_i$, $B_i$ and $C_i$. To do so, we first compute the derivative of $\mu_{ij}$ with respect to $\beta_{pd}$ for a given feature $p$ and a given class $d$. Let 
\[
E_i^\Sigma = \sum_{j'=1}^J \exp( \sum_{k=1}^K X_{ik}\beta_{kj'}).
\]
We consider the derivatives in two seperate cases.
\begin{itemize}
    \item {\bf Case $j \neq d$:}
    \begin{align}
        \frac{\partial \mu_{ij}}{\partial \beta_{pd}} &= \frac{-\exp(\sum_k X_{ik} \beta_{kj}) \times X_{ip} \exp(\sum_k X_{ik} \beta_{kd}) }
        {(E_i^\Sigma)^2} \notag\\
        &= - X_{ip} \mu_{ij} \mu_{id} 
        \label{eq:partial_mu_ij}
    \end{align}
    
    \item {\bf Case $j=d$:}
    \begin{align}
        \frac{\partial \mu_{id}}{\partial \beta_{pd}} &= \frac{ X_{ip} \exp(\sum_k X_{ik} \beta_{kd}) \times E_i^\Sigma
        -\exp(\sum_k X_{ik} \beta_{kd}) \times X_{ip} \exp(\sum_k X_{ik} \beta_{kd}) }
        {(E_i^\Sigma)^2} \notag\\
        &= X_{ip} \mu_{id} - X_{ip} (\mu_{id})^2\notag\\
        &= X_{ip} \mu_{id} (1 - \mu_{id}) 
        \label{eq:partial_mu_id}
    \end{align}
\end{itemize}
Since $\phi_i$ is not a function of $\beta$, for all $p$ and $d$, we have $\frac{\partial A_i}{\partial \beta_{pd}} = 0$ and 
\[
\frac{\partial \phi_i \mu_{ij}}{\partial \beta_{pd}} = \phi_i \frac{\partial \mu_{ij}}{\partial \beta_{pd}}.
\]
Now let $\psi$ be the digamma function, defined in \eqref{eq:digamma}. From \eqref{eq:partial_mu_ij} and \eqref{eq:partial_mu_id}, the derivatives of $B_i$ and $C_i$ are
\begin{align}
    \frac{\partial B_i}{\partial \beta_{pd}}  &= \phi X_{ip} \mu_{id} (1 - \mu_{id}) \cdot \frac{\partial}{\partial \phi_i \mu_{id}} \ln \big( \Gamma (\phi_i \mu_{id}) \big)
    - \phi_i \sum_{j \neq d} X_{ip} \mu_{ij} \mu_{id} \cdot \frac{\partial}{\partial \phi_i \mu_{ij}} \ln \big( \Gamma (\phi_i \mu_{ij}) \big) \notag\\
    &= \phi_i X_{ip} \mu_{id} \psi (\phi_i \mu_{id}) - \phi_i \sum_j X_{ip} \mu_{ij} \mu_{id} \psi (\phi_i \mu_{ij}) \notag\\
    &= \phi_i X_{ip} \mu_{id} \left(  \psi (\phi_i \mu_{id}) - \sum_j \mu_{ij} \psi (\phi_i \mu_{ij}) \right),\label{eq:b_i_mu}
\end{align}
and
\begin{align}
    \frac{\partial C_i}{\partial \beta_{pd}}  &= X_{ip} \mu_{id} (1 - \mu_{id}) \cdot \frac{\partial}{\partial \mu_{id}} (\phi_i \mu_{id}-1) \ln (y_{id})
    - \sum_{j \neq d} X_{ip} \mu_{ij} \mu_{id} \cdot \frac{\partial}{\partial \mu_{ij}} (\phi_i \mu_{ij}-1) \ln (y_{ij}) \notag\\
    &= \phi_i X_{ip} \mu_{id} \left(   \ln (y_{id}) - \sum_j \mu_{ij}  \ln (y_{ij}) \right). \label{eq:c_i_mu}
\end{align}
Summing the results of  \eqref{eq:b_i_mu} and \eqref{eq:c_i_mu} gives \eqref{eq:gradient}.

We now compute the derivatives of the loglikelihood with respect to $\gamma_k$ for $k \in [1,\dots,K_Z]$. Since
\[
\frac{\partial \phi_i}{\partial \gamma_k} = Z_{ik} \phi_i,
\]
using the change of variables,
\begin{align*}
    \frac{\partial A_i}{\partial \gamma_k}  &= Z_{ik} \phi_i \psi(\phi_i), \\
    \frac{\partial B_i}{\partial \gamma_k}  &=  \sum_j \psi(\phi_i \mu_{ij}) \mu_{ij} \frac{\partial \phi_i}{\partial \gamma_k} = Z_{ik} \phi_i \sum_j \mu_{ij} \psi(\phi_i \mu_{ij}), \quad \text{and}\\
    \frac{\partial C_i}{\partial \gamma_k}  &= Z_{ik} \phi_i \sum_j \mu_{ij} \ln (y_{ij}).
\end{align*}
Thus, the summation of the above three expression results as  \eqref{eq:derivative_gamma}.

\subsection{Second order derivative}
\label{sec:second-order-without}
The Hessian matrix of the loglikelihhod function \eqref{eqn:loglikelihood-again} is the collection of derivatives of \eqref{eq:gradient} with respect to all the variables in $\beta$. We divide the computation of the Hessian into two cases as shown below. Towards this, let $\psi_1$ be the trigamma function, which is the derivative of the digamma function. 
\begin{itemize}
    \item {\bf Case $c \neq d$:} In this case, using \eqref{eq:partial_mu_ij}, we have
\begin{equation}
        \frac{\partial}{\partial \beta_{qc}} X_{ip} \mu_{id} = - X_{iq} X_{ip} \mu_{ic} \mu_{id}.
\end{equation}
Furthermore, since
\[
\frac{\partial}{\partial \beta_{qc}} \sum_j \mu_{ij} \psi(\phi_i \mu_{ij}) = \sum_j \left[\psi(\phi_i \mu_{ij}) \frac{\partial \mu_{ij}}{\partial \beta_{qc}} + \mu_{ij} \frac{\partial \psi(\phi_i \mu_{ij})}{\partial \beta_{qc}}\right],
\]
using \eqref{eq:partial_mu_ij} and \eqref{eq:partial_mu_id},
\begin{align*}
        \frac{\partial}{\partial \beta_{qc}} \sum_j \mu_{ij} \psi(\phi_i \mu_{ij}) &= X_{iq} \mu_{ic} \cdot \frac{\partial}{\partial \mu_{ic}} \mu_{ic} \psi(\phi_i \mu_{ic}) - X_{iq} \mu_{ic}  \sum_j \mu_{ij} \cdot \frac{\partial}{\partial \mu_{ij}} \mu_{ij} \psi(\phi_i \mu_{ij}) \notag\\
         &= X_{iq} \mu_{ic} \bigg( \psi(\phi_i \mu_{ic}) + \phi_i \mu_{ic} \psi_1(\phi_i \mu_{ic}) - \sum_j \mu_{ij} \big( \psi(\phi_i \mu_{ij}) + \phi_i \mu_{ij} \psi_1(\phi_i \mu_{ij}) \big) \bigg),\\
        \frac{\partial}{\partial \beta_{qc}} \sum_j \mu_{ij} \ln(y_{ij}) &= X_{iq} \mu_{ic} \ln(y_{ic}) -  X_{iq} \mu_{ic}  \sum_j \mu_{ij} \ln(y_{ij}), 
\end{align*}
and
\begin{align*}
        \frac{\partial}{\partial \beta_{qc}} \big( \ln(y_{id}) - \psi(\phi_i \mu_{id}) \big) &= - \phi_i X_{iq} \mu_{id} \mu_{ic} \frac{\partial}{\partial \phi_i \mu_{id}} \big(\ln(y_{id}) - \psi(\phi_i \mu_{id}) \big) = \phi_i X_{iq} \mu_{id} \mu_{ic} \psi_1(\phi_i \mu_{id}). 
    \end{align*}
    By combining the above, we obtain:
    \begin{align}
        \frac{\partial}{\partial \beta_{qc}}  \frac{\partial}{\partial \beta_{pd}} \ell (y | \mu, \phi) &= \sum_{i=1}^n \bigg[ \phi_i X_{ip} \mu_{id} X_{iq} \mu_{ic} \bigg(
        \psi(\phi_i \mu_{ic}) + \phi_i \mu_{ic} \psi_1(\phi_i \mu_{ic}) - \sum_j \mu_{ij} \big( \psi(\phi_i \mu_{ij}) + \phi_i \mu_{ij} \psi_1(\phi_i \mu_{ij}) \big) \notag\\
        & \hspace{80pt} - \ln(y_{ic}) + \sum_j \mu_{ij} \ln(y_{ij})  + \phi_i \mu_{id} \psi_1(\phi_i \mu_{id}) \bigg) \notag\\
        & \hspace{30pt} - \phi_i X_{ip} \mu_{id} X_{iq} \mu_{ic} \bigg( \sum_j \mu_{ij} \big( \psi(\phi_i \mu_{ij}) - \ln(y_{ij}) \big) 
    - \psi(\phi_i \mu_{id}) + \ln(y_{id}) \bigg) \bigg] \notag \\
    &= \sum_{i=1}^n \bigg[ \phi_i X_{ip} \mu_{id} X_{iq} \mu_{ic} \bigg(
        \psi(\phi_i \mu_{ic}) + \phi_i \mu_{ic} \psi_1(\phi_i \mu_{ic}) - \phi_i \sum_j \mu_{ij}^ 2 \psi_1(\phi_i \mu_{ij}) \notag\\
        & \hspace{80pt} - \ln(y_{ic}) + 2 \sum_j \mu_{ij} \ln(y_{ij})  + \phi_i \mu_{id} \psi_1(\phi_i \mu_{id}) \notag\\
        & \hspace{80pt} - 2 \sum_j \mu_{ij} \psi(\phi_i \mu_{ij})
    + \psi(\phi_i \mu_{id}) - \ln(y_{id}) \bigg) \bigg]. \label{eq:hessian_beta_1}
    \end{align}
    
    \item {\bf Case $c=d$:}
    We now have
    \begin{equation}
        \frac{\partial}{\partial \beta_{qc}} X_{ip} \mu_{id} = X_{iq} \mu_{ic} (1 - \mu_{ic}) X_{ip},
    \end{equation}
    and
    \begin{equation}
        \frac{\partial}{\partial \beta_{qc}} \big( \ln(y_{ic}) - \psi(\phi_i \mu_{ic}) \big) = - \phi_i X_{iq} \mu_{ic} (1 - \mu_{ic}) \psi_1(\phi_i \mu_{ic}).
    \end{equation}  
    Thus, 
    \begin{align}
        \frac{\partial}{\partial \beta_{qc}}  \frac{\partial}{\partial \beta_{pc}} \ell (y | \mu, \phi) &= \sum_{i=1}^n \bigg[ \phi_i X_{ip} X_{iq} \mu_{ic}^2 \bigg(
        \psi(\phi_i \mu_{ic}) + \phi_i \mu_{ic} \psi_1(\phi_i \mu_{ic}) - \sum_j \mu_{ij} \big( \psi(\phi_i \mu_{ij}) + \phi_i \mu_{ij} \psi_1(\phi_i \mu_{ij}) \big) \notag\\
        & \hspace{80pt} - \ln(y_{ic}) + \sum_j \mu_{ij} \ln(y_{ij})  - \phi_i (1 - \mu_{ic}) \psi_1(\phi_i \mu_{ic}) \bigg) \notag\\
        & \hspace{30pt} + \phi_i X_{ip} X_{iq} \mu_{ic} (1 - \mu_{ic}) \bigg( \sum_j \mu_{ij} \big( \psi(\phi_i \mu_{ij}) - \ln(y_{ij}) \big) - \psi(\phi_i \mu_{ic}) + \ln(y_{ic}) \bigg) \bigg] \notag \\
    &= \sum_{i=1}^n \bigg[ \phi_i X_{ip} X_{iq} \mu_{ic}^2 \bigg(
        2 \psi(\phi_i \mu_{ic}) + 2 \phi_i \mu_{ic} \psi_1(\phi_i \mu_{ic}) - 2 \sum_j \mu_{ij} \psi(\phi_i \mu_{ij})  \notag\\
        & \hspace{80pt} - \phi_i \sum_j \mu_{ij}^2 \psi_1(\phi_i \mu_{ij}) - 2 \ln(y_{ic}) + 2 \sum_j \mu_{ij} \ln(y_{ij})  - \phi_i \psi_1(\phi_i \mu_{ic}) \bigg) \notag\\
        & \hspace{30pt} + \phi_i X_{ip} X_{iq} \mu_{ic} \bigg( \sum_j \mu_{ij} \big( \psi(\phi_i \mu_{ij}) - \ln(y_{ij}) \big) - \psi(\phi_i \mu_{ic}) + \ln(y_{ic}) \bigg) \bigg]. \label{eq:hessian_beta_2}
    \end{align}
\end{itemize}
Furthermore, we have:
\begin{align}
    \frac{\partial^2}{\partial \gamma_k \gamma_m} \ell (y | \mu, \phi) &= \sum_{i=1}^n \Bigg( \phi_i Z_{ik} \bigg( \phi_i Z_{im} \psi_1(\phi_i) - \phi_i \sum_j \mu_{ij} Z_{im} \mu_{ij} \psi_1(\phi_i \mu_{ij}) \bigg) \notag\\
    & \hspace{30pt} + \phi_i Z_{ik} Z_{im} \bigg( \psi(\phi_i) + \sum_j \mu_{ij} \big( \ln (y_{ij}) - \psi(\phi_i \mu_{ij}) \big) \bigg) \Bigg) \notag\\
    &= \sum_{i=1}^n \phi_i Z_{ik} Z_{im} \bigg( \phi_i \psi_1(\phi_i) + \psi(\phi_i)  - \phi_i \sum_j \mu_{ij}^2 \psi_1(\phi_i \mu_{ij}) + \sum_j \mu_{ij} \big( \ln (y_{ij}) - \psi(\phi_i \mu_{ij}) \big)  \bigg). \label{eq:second_derivative_gamma}
\end{align}
Finally, by differentiating \ref{eq:derivative_gamma}, we can compute the second derivative with respect to $\beta$ and $\gamma$,

\begin{align}
    \frac{\partial^2}{\partial \gamma_k \partial \beta_{pd}} \ell (y | \mu, \phi) &= \sum_{i=1}^n Z_{ik} \phi_i \frac{\partial}{\partial \beta_{pd}} \bigg(  \sum_j \mu_{ij} \big( \ln (y_{ij}) - \psi(\phi_i \mu_{ij}) \big)  \bigg) \notag\\
    &= \sum_{i=1}^n Z_{ik} \phi_i \bigg( X_{ip} \mu_{id} \frac{\partial}{\partial \mu_{id}} \mu_{id} \big( \ln (y_{id}) - \psi(\phi_i \mu_{id}) \big) \notag \\
    &\hspace{80pt}- \sum_j X_{ip} \mu_{id} \mu_{ij} \frac{\partial}{\partial \mu_{ij}} \mu_{ij} \big( \ln (y_{ij}) - \psi(\phi_i \mu_{ij}) \big)  \bigg) \notag\\
    &= \sum_{i=1}^n Z_{ik} \phi_i X_{ip} \mu_{id} \bigg( \big( \ln (y_{id}) - \psi(\phi_i \mu_{id}) - \phi_i \mu_{id} \psi_1(\phi_i \mu_{id}) \big) \notag\\
    & \hspace{80pt} - \sum_j \mu_{ij} \big(  \ln (y_{ij}) - \psi(\phi_i \mu_{ij}) - \phi_i \mu_{ij} \psi_1(\phi_i \mu_{ij}) \big)  \bigg).
\end{align}

\section{Computation on Dirichlet Distribution with Spatial Lag}
\label{appendix:with_spatial}

Given a matrix of coefficients $\beta \in \mathbb R^{K \times J}$, we take 
\[
E_i^\Sigma = \sum_{j'=1}^J \exp( \sum_{i'=1}^n \sum_{k=K}^n M_{ii'}^{-1} X_{i'k} \beta_{kj'}),
\]
and define $\mu$ as a matrix with elements 
\begin{equation*}
     \mu_{ij} = \frac{\exp( \sum_{i'=1}^n \sum_{k=1}^K M_{ii'}^{-1} X_{i'k} \beta_{kj})}{E_i^\Sigma} = \frac{\exp([M^{-1} X \beta]_{ij})}{\sum_{j'=1}^J \exp([M^{-1} X \beta]_{ij'})}.
\end{equation*}

In subsections \ref{sec:first-order-with} and \ref{sec:second-order-with}, we provide the computations of the first and second derivatives of the loglikelihood function with respect to $\beta$, $\gamma$ and $\rho$.

\subsection{First order derivative}
\label{sec:first-order-with}

Define the matrix $\tilde X \in \mathbb R^{n \times K}$ as  $\tilde X = M^{-1}X$. Then the derivative of the log-likelihood  with respect to $\beta$ consists of 
\begin{equation*}
    \frac{\partial}{\partial \beta_{pd}} \ell (y | \mu, \phi) = \sum_{i=1}^n \left( \phi_i \tilde X_{ip} \mu_{id} \bigg( \sum_j \mu_{ij} \big( \psi(\phi_i \mu_{ij}) - \ln(y_{ij}) \big) 
    - \psi(\phi_i \mu_{id}) + \ln(y_{id})
    \bigg)  \right).
\end{equation*}
We further differentiate the log-likelihood with respect to $\rho$. Note that 
\begin{align*} 
\frac{\partial M^{-1}}{\partial \rho}  &= - M^{-1} \frac{\partial M}{\partial \rho}   M^{-1} \\
&= M^{-1}  W  M^{-1}.
\end{align*}
Hence,
\begin{align*} 
\frac{\partial}{\partial \rho} (M^{-1}  X  \beta) &= \frac{\partial M^{-1}}{\partial \rho}   X  \beta \\
&= M^{-1}  W  M^{-1}  X  \beta.
\end{align*}
And thus, by defining $U = M^{-1}  W  M^{-1}  X  \beta$, we obtain
\begin{align} 
\frac{\partial \mu_{ij}}{\partial \rho} &= \frac{U_{ij} \,\exp([M^{-1} X \beta]_{ij}) \, \sum_{j'=1}^J \exp([M^{-1} X \beta]_{ij'}) -  \exp([M^{-1} X \beta]_{ij}) \, \sum_{j'=1}^J U_{ij'} \exp([M^{-1} X \beta]_{ij'}) }{\left( \sum_{j'=1}^J \exp([M^{-1} X \beta]_{ij'}) \right)^2} \notag \\
&= \mu_{ij} U_{ij}  - \mu_{ij} 
\sum_{j'=1}^J \mu_{ij'} U_{ij'}. \label{eq:der_mu_rho}
\end{align}
Using \eqref{eq:der_mu_rho}, we can then compute the derivative of the loglikelihood with respect to $\rho$ for each of the three elements defined in  \eqref{eqn:loglikelihood-again}. Then,
\begin{align*}
    \frac{\partial A_i}{\partial \rho}  &= 0, \\ 
    \frac{\partial B_i}{\partial \rho}  &= \sum_j \left( \phi_i \big( \mu_{ij} U_{ij}  - \mu_{ij} \sum_{j'=1}^J \mu_{ij'} U_{ij'} \big) \frac{\partial}{\partial \phi_i \mu_{ij}} \ln ( \Gamma(\phi_i \mu_{ij}) ) \right) \\
    &= \phi_i \sum_j \mu_{ij} U_{ij} \psi(\phi_i \mu_{ij}) -  \phi_i \sum_j \mu_{ij} U_{ij} \cdot \sum_j \mu_{ij} \psi(\phi_i \mu_{ij}) \\
    &= \phi_i \sum_j \mu_{ij} U_{ij} \Big( \psi(\phi_i \mu_{ij}) -   \sum_{j'} \mu_{ij'} \psi(\phi_i \mu_{ij'}) \Big), \\
    \frac{\partial C_i}{\partial \rho}  &= \sum_j \left( \big( \mu_{ij} U_{ij}  - \mu_{ij} \sum_{j'=1}^J \mu_{ij'} U_{ij'} \big) \frac{\partial}{\partial \mu_{ij}}(\phi_i \mu_{ij}-1) \ln (y_{ij}) \right) \\
    &= \phi_i \sum_j \mu_{ij} \ln (y_{ij}) \Big( U_{ij}  - \sum_{j'} \mu_{ij'} U_{ij'} \Big).
\end{align*}
By taking the summation of the above, we obtain \eqref{eq:derivative_ll_rho}.

\subsection{Second order derivative}
\label{sec:second-order-with}

Let 
\begin{equation}
\label{eq:Fij}
F_{ij} = \ln (y_{ij}) \big( U_{ij}  - \sum_{j'} \mu_{ij'} U_{ij'} \big),
\end{equation}
and 
\begin{equation}
\label{eq:Gij}
G_{ij} = U_{ij} \big( \psi(\phi_i \mu_{ij}) -   \sum_{j'} \mu_{ij'} \psi(\phi_i \mu_{ij'}) \big).
\end{equation}
Then,
\begin{align}
    \frac{\partial^2}{\partial \rho^2} \ell (y | \mu, \phi) &= \sum_{i=1}^n \phi_i \bigg( \sum_j \frac{\partial \mu_{ij}}{\partial \rho} \cdot (F_{ij} - G_{ij}) + \sum_j \mu_{ij} \cdot \Big( \frac{\partial}{\partial \rho} F_{ij} - \frac{\partial}{\partial \rho} G_{ij} \Big) \bigg). \label{eq:second_derivative_rho}
\end{align}
Observe that 
\begin{align}
    \frac{\partial U}{\partial \rho}  &= \frac{\partial}{\partial \rho} (M^{-1} W M^{-1} X \beta) \notag\\
    &= \frac{\partial}{\partial \rho} (M^{-1} W) M^{-1} X \beta + M^{-1} W  \frac{\partial}{\partial \rho} (M^{-1} X \beta) \notag\\
    &= 2  M^{-1} W M^{-1} W M^{-1} X \beta = 2 M^{-1} W U. \label{eq:derivative_U}
\end{align}
Furthermore, since
\begin{align}
    \frac{\partial}{\partial \rho} (\mu_{ij} U_{ij}) &= \mu_{ij} \frac{\partial}{\partial \rho} U_{ij} + U_{ij} \frac{\partial}{\partial \rho} \mu_{ij} \notag\\
    &= 2 \mu_{ij} V_{ij} + U_{ij} \big( \mu_{ij} U_{ij}  - \mu_{ij} \sum_{j'=1}^J \mu_{ij'} U_{ij'} \big), \label{eq:derivative_mu_U}
\end{align}
by taking $V = M^{-1} W U$, we observe that
\begin{align}
    \frac{\partial}{\partial \rho} \sum_j(\mu_{ij} U_{ij}) &= \sum_j \Big( 2 \mu_{ij} V_{ij} + U_{ij} \big( \mu_{ij} U_{ij}  - \mu_{ij} \sum_{j'} \mu_{ij'} U_{ij'} \big) \Big) \notag\\
    &= 2 \sum_j \big( \mu_{ij} V_{ij} \big) + \sum_j \big( \mu_{ij} U_{ij}^2 \big) - \sum_j \big( \mu_{ij} U_{ij} \sum_{j'} \mu_{ij'} U_{ij'} \big) \notag\\
    &= \sum_j \mu_{ij} \big( 2 V_{ij} + U_{ij}^2 \big) - \Big( \sum_j \mu_{ij} U_{ij} \Big)^2.
\end{align}
Using \eqref{eq:derivative_U} and \eqref{eq:derivative_mu_U}, for all $i,j$, 
\begin{align}
    \frac{\partial F_{ij}}{\partial \rho}  &= \ln (y_{ij}) \frac{\partial}{\partial \rho} \big( U_{ij}  - \sum_{j'} \mu_{ij'} U_{ij'} \big) \notag\\
    &= \ln (y_{ij}) \bigg( 2 V_{ij}  - \sum_{j'} \mu_{ij'} \big( 2 V_{ij'} + U_{ij'}^2 \big) + \Big( \sum_{j'} \mu_{ij'} U_{ij'} \Big)^2 \bigg), \label{eq:derivative_Fij}
\end{align}
and then using \eqref{eq:der_mu_rho},
\begin{align}
    \frac{\partial G_{ij}}{\partial \rho}  &= \frac{\partial U_{ij}}{\partial \rho}  \bigg( \psi(\phi_i \mu_{ij}) -   \sum_{j'} \mu_{ij'} \psi(\phi_i \mu_{ij'}) \bigg) + U_{ij} \, \frac{\partial}{\partial \rho} \bigg( \psi(\phi_i \mu_{ij}) -   \sum_{j'} \mu_{ij'} \psi(\phi_i \mu_{ij'}) \bigg) \notag\\
    &= 2 V_{ij} \bigg( \psi(\phi_i \mu_{ij}) -   \sum_{j'} \mu_{ij'} \psi(\phi_i \mu_{ij'}) \bigg) + U_{ij} \, \Big( \mu_{ij} U_{ij}  - \mu_{ij} \sum_{j'=1}^J \mu_{ij'} U_{ij'} \Big) \, \frac{\partial}{\partial \mu_{ij}} \psi(\phi_i \mu_{ij}) \notag\\
    & \hspace{20pt} - U_{ij} \,  \sum_{j'} \Big( \frac{\partial}{\partial \rho} \mu_{ij'} \psi(\phi_i \mu_{ij'}) \Big) \notag\\
    &= 2 V_{ij} \bigg( \psi(\phi_i \mu_{ij}) -   \sum_{j'} \mu_{ij'} \psi(\phi_i \mu_{ij'}) \bigg) + U_{ij} \phi_i \, \Big( \mu_{ij} U_{ij}  - \mu_{ij} \sum_{j'=1}^J \mu_{ij'} U_{ij'} \Big) \, \psi_1(\phi_i \mu_{ij}) \notag\\
    & \hspace{20pt} - U_{ij} \,  \sum_{j'} \bigg( \Big( \mu_{ij'} U_{ij'}  - \mu_{ij'} \sum_{k=1}^J \mu_{ik} U_{ik} \Big) \Big( \psi(\phi_i \mu_{ij'}) + \phi_i \mu_{ij'} \psi_1(\phi_i \mu_{ij'}) \Big) \bigg). \notag\\ \label{eq:derivative_Gij}
\end{align}

By introducing $\Omega_{ij} = \mu_{ij} U_{ij}$ and $\Omega_i^\Sigma = \sum_j \Omega_{ij}$, and noting $\alpha_{ij} = \phi_i \mu_{ij}$, we combine  \eqref{eq:derivative_Fij} and \eqref{eq:derivative_Gij} to get
\begin{align*}
    \frac{\partial}{\partial \rho} F_{ij} - \frac{\partial}{\partial \rho} G_{ij} &= \ln (y_{ij}) \bigg( 2 V_{ij}  - \sum_{j'} \mu_{ij'} \big( 2 V_{ij'} + U_{ij'}^2 \big) + (\Omega_i^\Sigma)^2 \bigg) - 2 V_{ij} \bigg( \psi(\alpha_{ij}) -   \sum_{j'} \mu_{ij'} \psi(\alpha_{ij'}) \bigg) \\
    & \hspace{20pt} - U_{ij} \phi_i \psi_1(\alpha_{ij}) \, \Big( \Omega_{ij}  - \mu_{ij} \Omega_i^\Sigma \Big) + U_{ij} \,  \sum_{j'} \bigg( \Big( \Omega_{ij}  - \mu_{ij} \Omega_i^\Sigma \Big) \Big( \psi(\alpha_{ij'}) + \alpha_{ij'} \psi_1(\alpha_{ij'}) \Big) \bigg).
\end{align*}
Inserting this expression into \eqref{eq:second_derivative_rho}, we can show that $    \frac{\partial^2}{\partial \rho^2} \ell (y | \mu, \phi)$ is equal to
\begin{align*}
 &\sum_{i=1}^n \phi_i \Bigg( \sum_j \big( \Omega_{ij}  - \mu_{ij} \Omega_i^\Sigma \big) \, (F_{ij} - G_{ij})  \\
    & \hspace{34pt} + \sum_j \mu_{ij} \bigg( \ln (y_{ij}) \Big( 2 V_{ij}  - \sum_{j'} \mu_{ij'} \big( 2 V_{ij'} + U_{ij'}^2 \big) + (\Omega_i^\Sigma)^2 \Big) - 2 V_{ij} \Big( \psi(\alpha_{ij}) -   \sum_{j'} \mu_{ij'} \psi(\alpha_{ij'}) \Big) \\
    & \hspace{70pt} - U_{ij} \phi_i \psi_1(\alpha_{ij}) \, \Big( \Omega_{ij}  - \mu_{ij} \Omega_i^\Sigma \Big) + U_{ij} \,  \sum_{j'} \bigg( \Big( \Omega_{ij}  - \mu_{ij} \Omega_i^\Sigma \Big) \Big( \psi(\alpha_{ij'}) + \alpha_{ij'} \psi_1(\alpha_{ij'}) \Big) \bigg) \Bigg). 
\end{align*}

Furthermore, differentiating \eqref{eq:derivative_ll_rho} with respect to $\gamma_k$ gives 
\begin{align}
    \frac{\partial^2}{\partial \rho \partial \gamma_k} \ell (y | \mu, \phi) &= \sum_{i=1}^n \Bigg( Z_{ik} \phi_i \sum_j \mu_{ij} \bigg( \ln (y_{ij}) \Big( U_{ij}  - \sum_{j'} \mu_{ij'} U_{ij'} \Big) - U_{ij} \Big( \psi(\phi_i \mu_{ij}) -   \sum_{j'} \mu_{ij'} \psi(\phi_i \mu_{ij'}) \Big) \bigg) \notag\\
    & \hspace{40pt} - \phi_i \sum_j \mu_{ij} U_{ij} \Big( Z_{ik} \phi_i \mu_{ij} \psi_1(\phi_i \mu_{ij}) -  \sum_{j'} Z_{ik} \phi_i \mu_{ij'}^2 \psi_1(\phi_i \mu_{ij'})  \Big)  \Bigg) \notag\\
    &= \sum_{i=1}^n Z_{ik} \phi_i \Bigg( \sum_j \mu_{ij} \bigg( \ln (y_{ij}) \Big( U_{ij}  - \sum_{j'} \mu_{ij'} U_{ij'} \Big) - U_{ij} \Big( \psi(\phi_i \mu_{ij}) -   \sum_{j'} \mu_{ij'} \psi(\phi_i \mu_{ij'}) \Big) \notag\\
    & \hspace{100pt} - \phi_i U_{ij} \Big( \mu_{ij} \psi_1(\phi_i \mu_{ij}) -  \sum_{j'} \mu_{ij'}^2 \psi_1(\phi_i \mu_{ij'}) \Big) \bigg) \notag\\
    &= \sum_{i=1}^n Z_{ik} \phi_i \Bigg( \sum_j \mu_{ij} \bigg( \ln (y_{ij}) \Big( U_{ij}  - \Omega_i^\Sigma \Big) - U_{ij} \Big( \psi(\alpha_{ij}) -   \sum_{j'} \mu_{ij'} \psi(\alpha_{ij'}) \Big) \notag\\
    & \hspace{100pt} - \phi_i U_{ij} \Big( \mu_{ij} \psi_1(\alpha_{ij}) -  \sum_{j'} \mu_{ij'}^2 \psi_1(\alpha_{ij'}) \Big) \bigg) \Bigg). \notag
\end{align}

Finally, to compute the derivative of \eqref{eq:derivative_ll_rho} with respect to $\beta_{pd}$, we need to compute the derivatives of $U_{ij}$ and $\sum_j \mu_{ij} U_{ij}$. First, we define $Q = M^{-1}  W  M^{-1}  X$.  Thus, $U=Q \beta$, that is, $U_{ij} = \sum_k Q_{ik} \beta_{kj}$. Hence, $\frac{\partial U_{ij}}{\partial \beta_{pd}} = Q_{ip}$ if $d=j$, and 0 otherwise. Hence, by adapting \eqref{eq:partial_mu_ij} and \eqref{eq:partial_mu_id} to $\tilde X$, we have
\begin{align}
    \frac{\partial}{\partial \beta_{pd}} \mu_{ij} U_{ij} =
    \left\{ 
    \begin{array}{l} 
    Q_{ip} \mu_{id} + U_{id} \tilde X_{ip} \mu_{id} (1 - \mu_{id}) \text{, if } d=j. \\
    - U_{ij} \tilde X_{ip} \mu_{ij} \mu_{id} \text{, if } d \neq j. \end{array}\right. 
\end{align}
\begin{align}
    \frac{\partial}{\partial \beta_{pd}} \sum_j \mu_{ij} U_{ij} &= \mu_{id} \Big( 
 Q_{ip} +\tilde X_{ip} \big( U_{id} - \sum_j \mu_{ij} U_{ij} \big) \Big) \notag\\
    &= \mu_{id} \Big( 
 Q_{ip} +\tilde X_{ip} \big( U_{id} - \Omega_i^\Sigma \big) \Big).
\end{align}

In order to be more concise, we keep the same notations of $F_{ij}$ and $G_{ij}$ defined in \eqref{eq:Fij} and \eqref{eq:Gij}.
\begin{align}
    \frac{\partial^2}{\partial \rho \partial \beta_{pd}} \ell (y | \mu, \phi) &= \sum_{i=1}^n \phi_i \bigg( \sum_j \frac{\partial \mu_{ij}}{\partial \beta_{pd}} \cdot (F_{ij} - G_{ij}) + \sum_j \mu_{ij} \cdot \Big( \frac{\partial}{\partial \beta_{pd}} F_{ij} - \frac{\partial}{\partial \beta_{pd}} G_{ij} \Big) \bigg) \notag\\
    &= \sum_{i=1}^n \phi_i \bigg(\tilde X_{ip} \mu_{id} \Big( (F_{id} - G_{id}) - \sum_j \mu_{ij} (F_{ij} - G_{ij}) \Big) \notag\\
    & \hspace{50pt} + \sum_j \mu_{ij} \frac{\partial}{\partial \beta_{pd}} F_{ij} - \sum_j \mu_{ij} \frac{\partial}{\partial \beta_{pd}} G_{ij} \bigg).
\end{align}
Then,
\begin{equation*}
    \sum_j \mu_{ij} \frac{\partial}{\partial \beta_{pd}} F_{ij} = \mu_{id} \ln(y_{id}) Q_{ip} -  \mu_{id} \Big( 
 Q_{ip} +\tilde X_{ip} \big( U_{id} - \Omega_i^\Sigma \big) \Big) \sum_j \mu_{ij} \ln(y_{ij}),
\end{equation*}
and
\begin{align*}
    \sum_j \mu_{ij} \frac{\partial}{\partial \beta_{pd}} G_{ij} &= \sum_j \mu_{ij} \Big( \psi(\phi_i \mu_{ij}) -   \sum_{j'} \mu_{ij'} \psi(\phi_i \mu_{ij'}) \Big) \frac{\partial}{\partial \beta_{pd}} U_{ij} \\
    &\hspace{40pt} + \sum_j \mu_{ij} U_{ij} \frac{\partial}{\partial \beta_{pd}} \big( \psi(\phi_i \mu_{ij}) -   \sum_{j'} \mu_{ij'} \psi(\phi_i \mu_{ij'}) \big) \\
    &= Q_{ip} \mu_{id} \Big( \psi(\phi_i \mu_{id}) -   \sum_{j'} \mu_{ij'} \psi(\phi_i \mu_{ij'}) \Big)  \\
    &\hspace{40pt} + \sum_j \mu_{ij} U_{ij} \frac{\partial}{\partial \beta_{pd}} \psi(\phi_i \mu_{ij}) -  \sum_j \mu_{ij} U_{ij} \sum_j \frac{\partial}{\partial \beta_{pd}} \mu_{ij} \psi(\phi_i \mu_{ij}) \\
    &= Q_{ip} \mu_{id} \Big( \psi(\phi_i \mu_{id}) -   \sum_j \mu_{ij} \psi(\phi_i \mu_{ij}) \Big) \\
    & \hspace{10pt} +\tilde X_{ip} \mu_{id} \phi_i \Big( \mu_{id} U_{id} \psi_1(\phi_i \mu_{id}) - \sum_j \mu_{ij}^2 U_{ij} \psi_1(\phi_i \mu_{ij}) \Big) \\
    & \hspace{10pt} - \sum_j (\mu_{ij} U_{ij}) \cdot\tilde X_{ip} \mu_{id} \Big( \psi(\phi_i \mu_{id}) + \phi_i \mu_{id} \psi_1(\phi_i \mu_{id}) - \sum_j \mu_{ij} \big( \psi(\phi_i \mu_{ij}) + \phi_i \mu_{ij} \psi_1(\phi_i \mu_{ij}) \big) \Big) \\
    &= \mu_{id} \Bigg( Q_{ip} \Big( \psi(\alpha_{id}) -   \sum_j \mu_{ij} \psi(\alpha_{ij}) \Big) +\tilde X_{ip} \phi_i \Big( \Omega_{id} \psi_1(\alpha_{id}) - \sum_j \mu_{ij}^2 U_{ij} \psi_1(\alpha_{ij}) \Big) \\
    & \hspace{30pt} -\tilde X_{ip} \Omega_i^\Sigma \Big( \psi(\alpha_{id}) + \alpha_{id} \psi_1(\alpha_{id}) - \sum_j \mu_{ij} \big( \psi(\alpha_{ij}) + \alpha_{ij} \psi_1(\alpha_{ij}) \big) \Big) \Bigg).
\end{align*}
Finally,
\begin{align}
    \frac{\partial^2}{\partial \rho \partial \beta_{pd}} \ell (y | \mu, \phi) &= \sum_{i=1}^n \phi_i \mu_{id} \Bigg(\tilde X_{ip} \Big( (F_{id} - G_{id}) - \sum_j \mu_{ij} (F_{ij} - G_{ij}) \Big) \notag\\
    & \hspace{40pt} + \ln(y_{id}) Q_{ip} -  \Big( 
 Q_{ip} +\tilde X_{ip} \big( U_{id} - \Omega_i^\Sigma \big) \Big) \sum_j \mu_{ij} \ln(y_{ij}) \notag\\ 
    & \hspace{40pt} - Q_{ip} \Big( \psi(\alpha_{id}) -   \sum_j \mu_{ij} \psi(\alpha_{ij}) \Big) -\tilde X_{ip} \phi_i \Big( \Omega_{id} \psi_1(\alpha_{id}) - \sum_j \mu_{ij}^2 U_{ij} \psi_1(\alpha_{ij}) \Big) \notag\\
    & \hspace{80pt} +\tilde X_{ip} \Omega_i^\Sigma \Big( \psi(\alpha_{id}) + \alpha_{id} \psi_1(\alpha_{id}) - \sum_j \mu_{ij} \big( \psi(\alpha_{ij}) + \alpha_{ij} \psi_1(\alpha_{ij}) \big) \Big) \Bigg).
\end{align}

\section{Equivalence between crossentropy and multinomial}
\label{appendix:equivalence_crossentropy_multinomial}

Let $J \in \mathbb N$ be the number of classes and $K \in \mathbb N$ the number of features. Let $Y_i = (Y_{i1},\dots,Y_{iJ}) \sim \text{Mult}(n_i, p_{i1}, \dots, p_{iJ})$, where $\sum_j p_{ij} = 1$ and $\sum_j Y_{ij} = n_i$ with $Y_{ij}$ non negative integers. The probability mass function is
\begin{equation*}
P[Y_{i1}=y_{i1},\dots,Y_{iJ}=y_{iJ}] = c(n_i) \prod_{j=1}^J p_{ij}^{y_{ij}}
\end{equation*}
where $c(n_i)$ is a term depending on $n_i$.

Now consider the multinomial logit model where we link the $p_j$ to some covariate $X$. Let's define $X_i$ the $K$-dimensional vector for the sample $i$. Then, for $\beta_j \in \mathbb R^K$, 

\begin{equation}
p_{ij} = \frac{\exp(X_i^T \beta_j)}{\sum_{j'} \exp(X_i^T \beta_{j'})}.
\label{eq:pij}
\end{equation}

Let $\theta = (\beta_1, \dots, \beta_J)$. The loglikelihood for a sample of $N$ observations is
\begin{align*}
\ell(\theta; y_1, \dots, y_n) &= \sum_{i=1}^N \ln(c(n_i)) +  \sum_{i=1}^N \sum_{j=1}^J  y_{ij}  \ln(p_{ij}) \\
&\propto \sum_{i=1}^N \sum_{j=1}^J  y_{ij}  \ln(p_{ij}),
\end{align*}
and, by noting $\tilde{y_{ij}} = \frac{y_{ij}}{n_i} \in [0,1]$,
\begin{align*}
\ell(\theta; y_1, \dots, y_n) &\propto \sum_{i=1}^N \sum_{j=1}^J  n_i \tilde{y_{ij}} \ln(p_{ij})\\
&\propto \sum_{i=1}^N n_i \sum_{j=1}^J \tilde{y_{ij}} \ln(p_{ij}).
\end{align*}

Then, in the case where all the samples have the same size $m$, we have $n_i = m$ for all $i$. From this we have that

\begin{align*}
\ell(\theta; y_1, \dots, y_n) &\propto m \sum_{i=1}^N \sum_{j=1}^J \tilde{y_{ij}} \ln(p_{ij})\\
&\propto \sum_{i=1}^N \sum_{j=1}^J \tilde{y_{ij}} \ln(p_{ij}) = -CE(\tilde y, p).
\end{align*}

\section{Results of the multinomial model on the synthetic dataset}
\label{appendix:multinomial}

Tables \ref{tab:scores_synthetic_01_ce}, \ref{tab:scores_synthetic_05_ce} and \ref{tab:scores_synthetic_09_ce} present the bias, standard deviation and mean squared error of the multinomial models using the cross-entropy minimization on the synthetic Dirichlet generated data described in Section \ref{subsec:synthetic}. Table \ref{table:scores_test_synthetic_ce} describes the mean performances of the evaluated parameters (with $n=1000$) on a test set.

\begin{table}[h]
\small
	\begin{center}
        \caption{Estimated bias (averaged over 
    $n=100$ replications) of the estimated parameters of the cross-entropy minimization in the context of Dirichlet generated data with $\rho=0.1$. Standard deviations are presented within parenthesis, and mean squared errors within square brackets.}
	\label{tab:scores_synthetic_01_ce}
		\begin{tabular}{|c|c|c|c|c|c|c|}
			\hline
			 &  \multicolumn{3}{c|}{Spatial} & \multicolumn{3}{c|}{Not spatial} \\
			\hline
			Parameter & n=50 & n=200 & n=1000 & n=50 & n=200 & n=1000 \\
			\hline
			\multirow{2}{*}{$\beta_{01}$} & 0.019 (0.088) & 0.002 (0.051) & 0.002 (0.022) & 0.02 (0.102) & 0.0 (0.058) & 0.002 (0.025) \\
			 & [0.008] & [0.003] & [0.0] &  [0.011] &  [0.003] &  [0.001] \\
			\hline
			\multirow{2}{*}{$\beta_{02}$} & 0.021 (0.108) & 0.004 (0.048) & 0.003 (0.022) & 0.037 (0.13) & 0.022 (0.056) & 0.016 (0.027) \\
			 & [0.012] & [0.002] & [0.0] &  [0.018] &  [0.004] &  [0.001] \\
			\hline
			\multirow{2}{*}{$\beta_{11}$} & -0.052 (0.117) & -0.02 (0.051) & -0.002 (0.023) & -0.052 (0.117) & -0.018 (0.051) & -0.001 (0.024) \\
			 & [0.016] & [0.003] & [0.001] &  [0.016] &  [0.003] &  [0.001] \\
			\hline
			\multirow{2}{*}{$\beta_{12}$} & 0.141 (0.127) & 0.043 (0.076) & 0.008 (0.035) & 0.145 (0.129) & 0.046 (0.077) & 0.013 (0.036) \\
			 & [0.036] & [0.008] & [0.001] &  [0.038] &  [0.008] &  [0.001] \\
			\hline
			\multirow{2}{*}{$\beta_{21}$} & 0.016 (0.12) & 0.008 (0.051) & -0.001 (0.02) & 0.016 (0.118) & 0.006 (0.05) & -0.002 (0.02) \\
			 & [0.015] & [0.003] & [0.0] &  [0.014] &  [0.003] &  [0.0] \\
			\hline
			\multirow{2}{*}{$\beta_{22}$} & 0.091 (0.126) & 0.031 (0.063) & 0.006 (0.031) & 0.091 (0.128) & 0.031 (0.065) & 0.01 (0.033) \\
			 & [0.024] & [0.005] & [0.001] &  [0.025] &  [0.005] &  [0.001] \\
			\hline
			\multirow{2}{*}{$\rho$} & -0.006 (0.068) & -0.003 (0.033) & 0.001 (0.013) & / & / & / \\
			 & [0.005] & [0.001] & [0.0] & / & / & / \\
			\hline
		\end{tabular}
	\end{center}
\end{table}

\begin{table}
\small
	\begin{center}
	\caption{Estimated bias (averaged over 
    $n=100$ replications) of the estimated parameters of the cross-entropy minimization in the context of Dirichlet generated data with $\rho=0.5$. Standard deviations are presented within parenthesis, and mean squared errors within square brackets.}
	\label{tab:scores_synthetic_05_ce}
		\begin{tabular}{|c|c|c|c|c|c|c|}
			\hline
			 &  \multicolumn{3}{c|}{Spatial} & \multicolumn{3}{c|}{Not spatial} \\
			\hline
			Parameter & n=50 & n=200 & n=1000 & n=50 & n=200 & n=1000 \\
			\hline
			\multirow{2}{*}{$\beta_{01}$} & -0.001 (0.058) & 0.006 (0.031) & 0.001 (0.013) & 0.013 (0.216) & 0.015 (0.115) & 0.022 (0.051) \\
			 & [0.003] & [0.001] & [0.0] &  [0.047] &  [0.013] &  [0.003] \\
			\hline
			\multirow{2}{*}{$\beta_{02}$} & -0.005 (0.062) & 0.007 (0.027) & 0.001 (0.013) & 0.123 (0.377) & 0.206 (0.195) & 0.169 (0.086) \\
			 & [0.004] & [0.001] & [0.0] &  [0.157] &  [0.08] &  [0.036] \\
			\hline
			\multirow{2}{*}{$\beta_{11}$} & -0.036 (0.104) & -0.011 (0.052) & -0.002 (0.026) & -0.026 (0.139) & 0.001 (0.069) & 0.007 (0.033) \\
			 & [0.012] & [0.003] & [0.001] &  [0.02] &  [0.005] &  [0.001] \\
			\hline
			\multirow{2}{*}{$\beta_{12}$} & 0.194 (0.121) & 0.055 (0.082) & 0.015 (0.036) & 0.314 (0.211) & 0.211 (0.112) & 0.204 (0.054) \\
			 & [0.052] & [0.01] & [0.002] &  [0.143] &  [0.057] &  [0.045] \\
			\hline
			\multirow{2}{*}{$\beta_{21}$} & 0.033 (0.081) & 0.009 (0.05) & 0.004 (0.023) & 0.017 (0.117) & -0.021 (0.066) & -0.023 (0.031) \\
			 & [0.008] & [0.003] & [0.001] &  [0.014] &  [0.005] &  [0.001] \\
			\hline
			\multirow{2}{*}{$\beta_{22}$} & 0.121 (0.114) & 0.036 (0.066) & 0.01 (0.03) & 0.174 (0.194) & 0.103 (0.102) & 0.11 (0.048) \\
			 & [0.028] & [0.006] & [0.001] &  [0.068] &  [0.021] &  [0.014] \\
			\hline
			\multirow{2}{*}{$\rho$} & -0.008 (0.036) & 0.0 (0.015) & 0.001 (0.007) & / & / & / \\
			 & [0.001] & [0.0] & [0.0] & / & / & / \\
			\hline
		\end{tabular}
	\end{center}
\end{table}

\begin{table}
\small
	\begin{center}
	\caption{Estimated bias (averaged over 
    $n=100$ replications) of the estimated parameters of the cross-entropy minimization in the context of Dirichlet generated data with $\rho=0.9$. Standard deviations are presented within parenthesis, and mean squared errors within square brackets.}
	\label{tab:scores_synthetic_09_ce}
		\begin{tabular}{|c|c|c|c|c|c|c|}
			\hline
			 &  \multicolumn{3}{c|}{Spatial} & \multicolumn{3}{c|}{Not spatial} \\
			\hline
			Parameter & n=50 & n=200 & n=1000 & n=50 & n=200 & n=1000 \\
			\hline
			\multirow{2}{*}{$\beta_{01}$} & 0.002 (0.063) & 0.0 (0.009) & 0.0 (0.004) & 0.143 (1.442) & 0.062 (0.73) & 0.179 (0.292) \\
			 & [0.004] & [0.0] & [0.0] &  [2.099] &  [0.537] &  [0.117] \\
			\hline
			\multirow{2}{*}{$\beta_{02}$} & -0.004 (0.088) & -0.003 (0.01) & -0.0 (0.004) & 0.696 (1.836) & 0.845 (0.743) & 0.726 (0.34) \\
			 & [0.008] & [0.0] & [0.0] &  [3.853] &  [1.267] &  [0.643] \\
			\hline
			\multirow{2}{*}{$\beta_{11}$} & -0.188 (0.316) & -0.029 (0.054) & -0.003 (0.023) & -0.22 (0.37) & -0.179 (0.155) & -0.218 (0.067) \\
			 & [0.135] & [0.004] & [0.001] &  [0.185] &  [0.056] &  [0.052] \\
			\hline
			\multirow{2}{*}{$\beta_{12}$} & 0.643 (0.355) & 0.164 (0.112) & 0.033 (0.038) & 1.109 (0.426) & 1.291 (0.199) & 1.32 (0.099) \\
			 & [0.54] & [0.039] & [0.003] &  [1.411] &  [1.705] &  [1.752] \\
			\hline
			\multirow{2}{*}{$\beta_{21}$} & 0.137 (0.229) & 0.031 (0.052) & 0.005 (0.021) & 0.103 (0.305) & 0.065 (0.144) & 0.079 (0.058) \\
			 & [0.071] & [0.004] & [0.0] &  [0.104] &  [0.025] &  [0.01] \\
			\hline
			\multirow{2}{*}{$\beta_{22}$} & 0.5 (0.328) & 0.118 (0.081) & 0.025 (0.029) & 0.76 (0.409) & 0.802 (0.162) & 0.855 (0.083) \\
			 & [0.358] & [0.021] & [0.001] &  [0.745] &  [0.669] &  [0.739] \\
			\hline
			\multirow{2}{*}{$\rho$} & -0.004 (0.02) & 0.0 (0.003) & -0.0 (0.001) & / & / & / \\
			 & [0.0] & [0.0] & [0.0] & / & / & / \\
			\hline
		\end{tabular}
	\end{center}
\end{table}

\begin{table}[h!]
  \centering
  \caption{Performance measures on the test set of the Dirichlet synthetic dataset comparing $\mu^*$ and the estimated $\hat \mu$ (computed with the parameters estimated with $n=1000$ with the minimization of cross-entropy). The results are displayed as mean on the 100 iterations (standard deviation).}
  \label{table:scores_test_synthetic_ce}
  \begin{tabular}{|c|l|c|c|c|c|}
    \hline
    & \textbf{Model} & $\boldsymbol{R^2}$ & \textbf{RMSE} & \textbf{Cross-entropy} & \textbf{Cos similarity} \\
    \hline
    \multirow{2}{*}{$\rho=0.1$} & Not spatial & 0.9314 ($<10^{-4}$) & 0.0736 ($<10^{-4}$) & 0.6655 (0.0001) & 0.9856 ($<10^{-4}$) \\
    & Spatial & 0.9338 ($<10^{-4}$) & 0.072 ($<10^{-4}$)  & 0.6644 (0.0001) & 0.9862 ($<10^{-4}$) \\ \hline
    \multirow{2}{*}{$\rho=0.5$} & Not spatial & 0.8421 (0.0001) & 0.1207 ($<10^{-4}$) & 0.6764 (0.0002) & 0.9618 ($<10^{-4}$)\\
    & Spatial & 0.9414 ($<10^{-4}$) & 0.07 ($<10^{-4}$) & 0.627 (0.0002) & 0.9872 ($<10^{-4}$) \\ \hline
    \multirow{2}{*}{$\rho=0.9$} & Not spatial & 0.274 (0.0021) & 0.3121 (0.0002) & 0.8576 (0.0019) & 0.7877 (0.0003)  \\
    & Spatial & 0.9761 ($<10^{-4}$) & 0.0535 ($<10^{-4}$) & 0.3716 (0.0009) & 0.9933 ($<10^{-4}$) \\
    \hline
  \end{tabular}
\end{table}

\section{Results of the models on the synthetic multinomial data}
\label{appendix:synthetic_multinomial}

Here, the data are generated using a multinomial distribution. As in Section \ref{subsec:synthetic}, a value of $\beta$ is fixed. Here, we choose
\[
\beta = {\begin{bmatrix} 0 & -0.2 & 0.1 \\ 0 & 2 & -1.5 \\ 0 & 0.5 & -2
\end{bmatrix}}.
\]
Here, the choice of $\beta$ differs from that in Section \ref{subsec:synthetic}. Indeed, using the $\beta$ from Section \ref{subsec:synthetic} to generate data from a multinomial distribution did not bring enough variation in the sample and resulted in similar outputs (i.e., the label vectors $y_i$ were identical for all $i$).

Once $\beta$ is chosen, for each $i \in [1,n]$, with $n$ the number of generated points (50, 200, or 1000), the number of trials $n_i$ is created by uniformly drawing a value between 100 and 10000. This value is fixed for all the 100 replications. 

The bias, standard deviation and mean squared error of the parameters estimated by the Dirichlet model on this synthetic dataset are presented in Tables \ref{tab:synthetic_multinomial_01}, \ref{tab:synthetic_multinomial_05} and \ref{tab:synthetic_multinomial_09}, and the results of the multinomial model are presented in Table \ref{tab:synthetic_multinomial_01_ce}, \ref{tab:synthetic_multinomial_05_ce} and \ref{tab:synthetic_multinomial_09_ce}. Tables \ref{table:scores_test_synthetic_multinomial} and \ref{table:scores_test_synthetic_multinomial_ce} present the performances measured on different metrics between the true $\mu^*$ of the test set and the estimated $\hat\mu$ computed with the parameters estimated with the $n=1000$ case (the same computations as in Section \ref{subsec:synthetic}).

The expected behaviour is observed: for both the Dirichlet and the multinomial models, the spatial lag model outperforms the non-spatial one, with a difference which is stronger when the strength of the spatial correlation ($\rho$) is higher. Furthermore, only under a high spatial correlation ($\rho=0.9$), the spatial multinomial model performs better than the spatial Dirichlet model, suggesting that both models are equivalent under a weak spatial correlation.

\begin{table}[h]
    \caption{Estimated bias (averaged over 
    $n=100$ replications) of the estimated parameters of the Dirichlet model in the context of   multinomial generated data with $\rho=0.1$. Standard deviations are presented within parenthesis, and mean squared errors within square brackets.}
    \label{tab:synthetic_multinomial_01}
	\begin{center}
		\begin{tabular}{|c|c|c|c|c|c|c|}
			\hline
			 &  \multicolumn{3}{c|}{Spatial} & \multicolumn{3}{c|}{Not spatial} \\
			\hline
			Parameter & n=50 & n=200 & n=1000 & n=50 & n=200 & n=1000 \\
			\hline
			\multirow{2}{*}{$\beta_{01}$} & 0.051 (0.032) & 0.016 (0.01) & 0.004 (0.003) & 0.043 (0.038) & 0.001 (0.019) & -0.008 (0.008) \\
			 & [0.004] & [0.0] & [0.0] &  [0.003] &  [0.0] &  [0.0] \\
			\hline
			\multirow{2}{*}{$\beta_{02}$} & 0.029 (0.025) & 0.01 (0.006) & 0.003 (0.002) & 0.047 (0.05) & 0.031 (0.022) & 0.021 (0.01) \\
			 & [0.001] & [0.0] & [0.0] &  [0.005] &  [0.001] &  [0.001] \\
			\hline
			\multirow{2}{*}{$\beta_{11}$} & -0.136 (0.046) & -0.039 (0.009) & -0.009 (0.004) & -0.15 (0.041) & -0.053 (0.016) & -0.027 (0.008) \\
			 & [0.021] & [0.002] & [0.0] &  [0.024] &  [0.003] &  [0.001] \\
			\hline
			\multirow{2}{*}{$\beta_{12}$} & 0.143 (0.037) & 0.04 (0.012) & 0.011 (0.004) & 0.162 (0.042) & 0.057 (0.017) & 0.033 (0.008) \\
			 & [0.022] & [0.002] & [0.0] &  [0.028] &  [0.004] &  [0.001] \\
			\hline
			\multirow{2}{*}{$\beta_{21}$} & -0.052 (0.027) & -0.016 (0.008) & -0.004 (0.003) & -0.064 (0.027) & -0.023 (0.011) & -0.014 (0.005) \\
			 & [0.003] & [0.0] & [0.0] &  [0.005] &  [0.001] &  [0.0] \\
			\hline
			\multirow{2}{*}{$\beta_{22}$} & 0.155 (0.046) & 0.044 (0.021) & 0.01 (0.004) & 0.169 (0.043) & 0.059 (0.018) & 0.032 (0.009) \\
			 & [0.026] & [0.002] & [0.0] &  [0.03] &  [0.004] &  [0.001] \\
			\hline
			\multirow{2}{*}{$\gamma_0$} & 6.439 (0.624) & 7.335 (0.471) & 7.78 (0.178) & 5.975 (0.373) & 6.274 (0.246) & 6.38 (0.125) \\
			 & [41.849] & [54.025] & [60.56] &  [35.845] &  [39.42] &  [40.714] \\
			\hline
			\multirow{2}{*}{$\gamma_1$} & 0.147 (0.803) & 0.152 (0.534) & 0.124 (0.306) & 0.093 (0.611) & 0.171 (0.427) & 0.101 (0.185) \\
			 & [0.667] & [0.308] & [0.109] &  [0.382] &  [0.212] &  [0.045] \\
			\hline
			\multirow{2}{*}{$\rho$} & -0.003 (0.022) & -0.001 (0.008) & -0.0 (0.001) & / & / & / \\
			 & [0.0] & [0.0] & [0.0] & / & / & / \\
			\hline
		\end{tabular}
	\end{center}
\end{table}

\begin{table}
    \caption{Estimated bias (averaged over 
    $n=100$ replications) of the estimated parameters of the Dirichlet model in the context of   multinomial generated data with $\rho=0.5$. Standard deviations are presented within parenthesis, and mean squared errors within square brackets.}
    \label{tab:synthetic_multinomial_05}
	\begin{center}
		\begin{tabular}{|c|c|c|c|c|c|c|}
			\hline
			 &  \multicolumn{3}{c|}{Spatial} & \multicolumn{3}{c|}{Not spatial} \\
			\hline
			Parameter & n=50 & n=200 & n=1000 & n=50 & n=200 & n=1000 \\
			\hline
			\multirow{2}{*}{$\beta_{01}$} & 0.041 (0.033) & 0.015 (0.02) & 0.007 (0.019) & 0.191 (0.22) & 0.157 (0.109) & 0.161 (0.044) \\
			 & [0.003] & [0.001] & [0.0] &  [0.085] &  [0.036] &  [0.028] \\
			\hline
			\multirow{2}{*}{$\beta_{02}$} & 0.016 (0.025) & 0.015 (0.046) & 0.01 (0.037) & 0.227 (0.264) & 0.271 (0.121) & 0.241 (0.057) \\
			 & [0.001] & [0.002] & [0.001] &  [0.121] &  [0.088] &  [0.061] \\
			\hline
			\multirow{2}{*}{$\beta_{11}$} & -0.151 (0.033) & -0.056 (0.077) & -0.033 (0.116) & -0.616 (0.194) & -0.62 (0.103) & -0.613 (0.054) \\
			 & [0.024] & [0.009] & [0.015] &  [0.417] &  [0.395] &  [0.378] \\
			\hline
			\multirow{2}{*}{$\beta_{12}$} & 0.163 (0.043) & 0.051 (0.042) & 0.023 (0.065) & 0.611 (0.177) & 0.592 (0.082) & 0.595 (0.043) \\
			 & [0.028] & [0.004] & [0.005] &  [0.405] &  [0.357] &  [0.356] \\
			\hline
			\multirow{2}{*}{$\beta_{21}$} & -0.062 (0.026) & -0.013 (0.025) & -0.001 (0.02) & -0.287 (0.104) & -0.263 (0.048) & -0.262 (0.021) \\
			 & [0.005] & [0.001] & [0.0] &  [0.093] &  [0.072] &  [0.069] \\
			\hline
			\multirow{2}{*}{$\beta_{22}$} & 0.172 (0.045) & 0.063 (0.08) & 0.038 (0.133) & 0.634 (0.19) & 0.624 (0.113) & 0.625 (0.05) \\
			 & [0.032] & [0.01] & [0.019] &  [0.438] &  [0.402] &  [0.393] \\
			\hline
			\multirow{2}{*}{$\gamma_0$} & 6.29 (0.482) & 7.177 (1.023) & 7.492 (1.23) & 2.827 (0.57) & 2.576 (0.296) & 2.496 (0.143) \\
			 & [39.8] & [52.556] & [57.639] &  [8.315] &  [6.725] &  [6.251] \\
			\hline
			\multirow{2}{*}{$\gamma_1$} & 0.068 (0.619) & 0.192 (0.509) & 0.178 (0.451) & 0.24 (0.783) & 0.146 (0.27) & 0.11 (0.139) \\
			 & [0.388] & [0.296] & [0.235] &  [0.67] &  [0.094] &  [0.031] \\
			\hline
			\multirow{2}{*}{$\rho$} & -0.001 (0.01) & -0.004 (0.022) & -0.004 (0.02) & / & / & / \\
			 & [0.0] & [0.0] & [0.0] & / & / & / \\
			\hline
		\end{tabular}
	\end{center}
\end{table}

\begin{table}
    \caption{Estimated bias (averaged over 
    $n=100$ replications) of the estimated parameters of the Dirichlet model in the context of multinomial generated data with $\rho=0.9$. Standard deviations are presented within parenthesis, and mean squared errors within square brackets.}
    \label{tab:synthetic_multinomial_09}
	\begin{center}
		\begin{tabular}{|c|c|c|c|c|c|c|}
			\hline
			 &  \multicolumn{3}{c|}{Spatial} & \multicolumn{3}{c|}{Not spatial} \\
			\hline
			Parameter & n=50 & n=200 & n=1000 & n=50 & n=200 & n=1000 \\
			\hline
			\multirow{2}{*}{$\beta_{01}$} & 0.117 (0.095) & 0.131 (0.093) & 0.12 (0.085) & 0.361 (0.651) & 0.18 (0.219) & 0.169 (0.084) \\
			 & [0.023] & [0.026] & [0.022] &  [0.555] &  [0.08] &  [0.036] \\
			\hline
			\multirow{2}{*}{$\beta_{02}$} & 0.023 (0.379) & 0.001 (0.056) & -0.009 (0.021) & 0.548 (0.948) & 0.42 (0.334) & 0.311 (0.146) \\
			 & [0.144] & [0.003] & [0.001] &  [1.199] &  [0.288] &  [0.118] \\
			\hline
			\multirow{2}{*}{$\beta_{11}$} & -1.097 (0.705) & -0.938 (0.665) & -0.839 (0.591) & -1.61 (0.29) & -1.639 (0.137) & -1.606 (0.063) \\
			 & [1.7] & [1.323] & [1.053] &  [2.677] &  [2.705] &  [2.583] \\
			\hline
			\multirow{2}{*}{$\beta_{12}$} & 0.847 (0.511) & 0.763 (0.522) & 0.714 (0.496) & 1.235 (0.23) & 1.215 (0.076) & 1.196 (0.042) \\
			 & [0.979] & [0.855] & [0.757] &  [1.577] &  [1.482] &  [1.432] \\
			\hline
			\multirow{2}{*}{$\beta_{21}$} & -0.441 (0.401) & -0.321 (0.226) & -0.295 (0.206) & -0.549 (0.205) & -0.526 (0.089) & -0.511 (0.035) \\
			 & [0.356] & [0.154] & [0.13] &  [0.344] &  [0.284] &  [0.262] \\
			\hline
			\multirow{2}{*}{$\beta_{22}$} & 0.95 (0.559) & 0.902 (0.633) & 0.83 (0.582) & 1.501 (0.26) & 1.496 (0.118) & 1.498 (0.049) \\
			 & [1.216] & [1.214] & [1.028] &  [2.322] &  [2.251] &  [2.246] \\
			\hline
			\multirow{2}{*}{$\gamma_0$} & 3.687 (1.675) & 3.561 (2.612) & 3.848 (2.86) & 1.025 (1.068) & 0.259 (0.205) & 0.041 (0.089) \\
			 & [16.397] & [19.503] & [22.985] &  [2.19] &  [0.109] &  [0.01] \\
			\hline
			\multirow{2}{*}{$\gamma_1$} & 0.182 (0.766) & 0.12 (0.181) & 0.144 (0.177) & 0.105 (0.555) & 0.126 (0.136) & 0.104 (0.061) \\
			 & [0.62] & [0.047] & [0.052] &  [0.319] &  [0.035] &  [0.015] \\
			\hline
			\multirow{2}{*}{$\rho$} & -0.015 (0.08) & -0.01 (0.054) & -0.003 (0.023) & / & / & / \\
			 & [0.007] & [0.003] & [0.001] & / & / & / \\
			\hline
		\end{tabular}
	\end{center}
\end{table}

\begin{table}[h!]
  \centering
  \caption{Performance measures on the test set of the multinomial synthetic dataset comparing $\mu^*$ and the estimated $\hat \mu$ (computed with the parameters estimated with $n=1000$ with the maximization of the Dirichlet distribution likelihood). The results are displayed as mean on the 100 iterations (standard deviation).}
  \label{table:scores_test_synthetic_multinomial}
  \begin{tabular}{|c|l|c|c|c|c|}
    \hline
    & \textbf{Model} & $\boldsymbol{R^2}$ & \textbf{RMSE} & \textbf{Cross-entropy} & \textbf{Cos similarity} \\
    \hline
    \multirow{2}{*}{$\rho=0.1$} & Not spatial & 0.9955 ($<10^{-4}$) & 0.0181 ($<10^{-4}$) & 0.5881 (0.0001) & 0.9990 ($<10^{-4}$) \\
    & Spatial & 0.9990 ($<10^{-4}$) & 0.0072 ($<10^{-4}$)  & 0.5866 (0.0001) & 0.9999 ($<10^{-4}$) \\ \hline
    \multirow{2}{*}{$\rho=0.5$} & Not spatial & 0.8400 (0.0001) & 0.1138 ($<10^{-4}$) & 0.6303 (0.0001) & 0.9669 ($<10^{-4}$)\\
    & Spatial & 0.9958 (0.0002) & 0.0098 (0.0001) & 0.5565 (0.0002) & 0.9994 ($<10^{-4}$) \\ \hline
    \multirow{2}{*}{$\rho=0.9$} & Not spatial & 0.1207 (0.0063) & 0.3245 (0.0001) & 0.8872 (0.0008) & 0.7777 (0.0002)  \\
    & Spatial & 0.8891 (0.0093) & 0.0870 (0.0039) & 0.4108 (0.0057) & 0.9784 (0.0004) \\
    \hline
  \end{tabular}
\end{table}

\begin{table}
    \caption{Estimated bias (averaged over 
    $n=100$ replications) of the estimated parameters of the multinomial model (via cross-entropy minimization) in the context of multinomial generated data with $\rho=0.1$. Standard deviations are presented within parenthesis, and mean squared errors within square brackets.}
    \label{tab:synthetic_multinomial_01_ce}
	\begin{center}
		\begin{tabular}{|c|c|c|c|c|c|c|}
			\hline
			 &  \multicolumn{3}{c|}{Spatial} & \multicolumn{3}{c|}{Not spatial} \\
			\hline
			Parameter & n=50 & n=200 & n=1000 & n=50 & n=200 & n=1000 \\
			\hline
			\multirow{2}{*}{$\beta_{01}$} & 0.067 (0.019) & 0.018 (0.006) & 0.003 (0.002) & 0.057 (0.038) & -0.001 (0.019) & -0.014 (0.008) \\
			 & [0.005] & [0.0] & [0.0] &  [0.005] &  [0.0] &  [0.0] \\
			\hline
			\multirow{2}{*}{$\beta_{02}$} & 0.043 (0.019) & 0.012 (0.006) & 0.003 (0.002) & 0.059 (0.05) & 0.03 (0.022) & 0.017 (0.01) \\
			 & [0.002] & [0.0] & [0.0] &  [0.006] &  [0.001] &  [0.0] \\
			\hline
			\multirow{2}{*}{$\beta_{11}$} & -0.159 (0.03) & -0.043 (0.008) & -0.008 (0.004) & -0.163 (0.041) & -0.043 (0.017) & -0.011 (0.007) \\
			 & [0.026] & [0.002] & [0.0] &  [0.028] &  [0.002] &  [0.0] \\
			\hline
			\multirow{2}{*}{$\beta_{12}$} & 0.176 (0.037) & 0.046 (0.01) & 0.01 (0.004) & 0.181 (0.045) & 0.048 (0.017) & 0.017 (0.008) \\
			 & [0.032] & [0.002] & [0.0] &  [0.035] &  [0.003] &  [0.0] \\
			\hline
			\multirow{2}{*}{$\beta_{21}$} & -0.073 (0.019) & -0.019 (0.006) & -0.004 (0.003) & -0.076 (0.027) & -0.02 (0.011) & -0.007 (0.005) \\
			 & [0.006] & [0.0] & [0.0] &  [0.006] &  [0.001] &  [0.0] \\
			\hline
			\multirow{2}{*}{$\beta_{22}$} & 0.177 (0.034) & 0.046 (0.009) & 0.01 (0.004) & 0.18 (0.043) & 0.048 (0.018) & 0.015 (0.008) \\
			 & [0.033] & [0.002] & [0.0] &  [0.034] &  [0.003] &  [0.0] \\
			\hline
			\multirow{2}{*}{$\rho$} & -0.001 (0.01) & -0.0 (0.004) & -0.0 (0.001) & / & / & / \\
			 & [0.0] & [0.0] & [0.0] & / & / & / \\
			\hline
		\end{tabular}
	\end{center}
\end{table}

\begin{table}
    \caption{Estimated bias (averaged over 
    $n=100$ replications) of the estimated parameters of the multinomial model (via cross-entropy minimization) in the context of multinomial generated data with $\rho=0.5$. Standard deviations are presented within parenthesis, and mean squared errors within square brackets.}
    \label{tab:synthetic_multinomial_05_ce}
	\begin{center}
		\begin{tabular}{|c|c|c|c|c|c|c|}
			\hline
			 &  \multicolumn{3}{c|}{Spatial} & \multicolumn{3}{c|}{Not spatial} \\
			\hline
			Parameter & n=50 & n=200 & n=1000 & n=50 & n=200 & n=1000 \\
			\hline
			\multirow{2}{*}{$\beta_{01}$} & 0.052 (0.014) & 0.013 (0.004) & 0.003 (0.001) & 0.076 (0.265) & -0.005 (0.143) & 0.015 (0.058) \\
			 & [0.003] & [0.0] & [0.0] &  [0.076] &  [0.02] &  [0.004] \\
			\hline
			\multirow{2}{*}{$\beta_{02}$} & 0.022 (0.015) & 0.007 (0.004) & 0.001 (0.001) & 0.219 (0.321) & 0.257 (0.161) & 0.232 (0.074) \\
			 & [0.001] & [0.0] & [0.0] &  [0.151] &  [0.092] &  [0.059] \\
			\hline
			\multirow{2}{*}{$\beta_{11}$} & -0.182 (0.037) & -0.047 (0.008) & -0.009 (0.004) & -0.278 (0.142) & -0.171 (0.087) & -0.157 (0.038) \\
			 & [0.035] & [0.002] & [0.0] &  [0.098] &  [0.037] &  [0.026] \\
			\hline
			\multirow{2}{*}{$\beta_{12}$} & 0.2 (0.049) & 0.052 (0.01) & 0.01 (0.003) & 0.34 (0.168) & 0.242 (0.081) & 0.238 (0.044) \\
			 & [0.042] & [0.003] & [0.0] &  [0.144] &  [0.065] &  [0.058] \\
			\hline
			\multirow{2}{*}{$\beta_{21}$} & -0.084 (0.026) & -0.022 (0.007) & -0.004 (0.003) & -0.174 (0.122) & -0.117 (0.056) & -0.124 (0.026) \\
			 & [0.008] & [0.001] & [0.0] &  [0.045] &  [0.017] &  [0.016] \\
			\hline
			\multirow{2}{*}{$\beta_{22}$} & 0.2 (0.044) & 0.054 (0.01) & 0.01 (0.003) & 0.285 (0.169) & 0.192 (0.088) & 0.185 (0.042) \\
			 & [0.042] & [0.003] & [0.0] &  [0.11] &  [0.045] &  [0.036] \\
			\hline
			\multirow{2}{*}{$\rho$} & -0.0 (0.008) & -0.0 (0.002) & 0.0 (0.001) & / & / & / \\
			 & [0.0] & [0.0] & [0.0] & / & / & / \\
			\hline
		\end{tabular}
	\end{center}
\end{table}

\begin{table}
    \caption{Estimated bias (averaged over 
    $n=100$ replications) of the estimated parameters of the multinomial model (via cross-entropy minimization) in the context of multinomial generated data with $\rho=0.9$. Standard deviations are presented within parenthesis, and mean squared errors within square brackets.}
    \label{tab:synthetic_multinomial_09_ce}
	\begin{center}
		\begin{tabular}{|c|c|c|c|c|c|c|}
			\hline
			 &  \multicolumn{3}{c|}{Spatial} & \multicolumn{3}{c|}{Not spatial} \\
			\hline
			Parameter & n=50 & n=200 & n=1000 & n=50 & n=200 & n=1000 \\
			\hline
			\multirow{2}{*}{$\beta_{01}$} & 0.072 (0.053) & 0.023 (0.01) & 0.004 (0.001) & 0.405 (1.403) & 0.367 (0.698) & 0.496 (0.245) \\
			 & [0.008] & [0.001] & [0.0] &  [2.132] &  [0.622] &  [0.307] \\
			\hline
			\multirow{2}{*}{$\beta_{02}$} & 0.006 (0.114) & -0.003 (0.003) & -0.001 (0.0) & 0.954 (1.45) & 1.113 (0.652) & 1.001 (0.289) \\
			 & [0.013] & [0.0] & [0.0] &  [3.013] &  [1.663] &  [1.087] \\
			\hline
			\multirow{2}{*}{$\beta_{11}$} & -0.637 (0.511) & -0.157 (0.067) & -0.029 (0.007) & -0.99 (0.448) & -0.99 (0.239) & -1.028 (0.081) \\
			 & [0.666] & [0.029] & [0.001] &  [1.181] &  [1.036] &  [1.064] \\
			\hline
			\multirow{2}{*}{$\beta_{12}$} & 0.545 (0.318) & 0.161 (0.065) & 0.031 (0.007) & 0.982 (0.363) & 1.136 (0.218) & 1.176 (0.1) \\
			 & [0.398] & [0.03] & [0.001] &  [1.096] &  [1.339] &  [1.393] \\
			\hline
			\multirow{2}{*}{$\beta_{21}$} & -0.295 (0.314) & -0.067 (0.029) & -0.012 (0.004) & -0.546 (0.324) & -0.627 (0.202) & -0.676 (0.083) \\
			 & [0.186] & [0.005] & [0.0] &  [0.403] &  [0.433] &  [0.464] \\
			\hline
			\multirow{2}{*}{$\beta_{22}$} & 0.562 (0.324) & 0.17 (0.067) & 0.033 (0.007) & 0.883 (0.425) & 0.94 (0.197) & 0.986 (0.082) \\
			 & [0.42] & [0.033] & [0.001] &  [0.961] &  [0.922] &  [0.979] \\
			\hline
			\multirow{2}{*}{$\rho$} & -0.009 (0.022) & -0.0 (0.001) & -0.0 (0.0) & / & / & / \\
			 & [0.001] & [0.0] & [0.0] & / & / & / \\
			\hline
		\end{tabular}
	\end{center}
\end{table}

\begin{table}[h!]
  \centering
  \caption{Performance measure on the test set of the multinomial synthetic dataset comparing $\mu^*$ and the estimated $\hat \mu$  (computed with the multinomial model with $n=1000$ via the minimization of cross-entropy). The results are displayed as mean on the 100 iterations (standard deviation).}
  \label{table:scores_test_synthetic_multinomial_ce}
  \begin{tabular}{|c|l|c|c|c|c|}
    \hline
    & \textbf{Model} & $\boldsymbol{R^2}$ & \textbf{RMSE} & \textbf{Cross-entropy} & \textbf{Cos similarity} \\
    \hline
    \multirow{2}{*}{$\rho=0.1$} & Not spatial & 0.9955 ($<10^{-4}$) & 0.0179 ($<10^{-4}$) & 0.5880 (0.0001) & 0.9991 ($<10^{-4}$) \\
    & Spatial & 0.9990 ($<10^{-4}$) & 0.0071 ($<10^{-4}$)  & 0.5866 (0.0001) & 0.9999 ($<10^{-4}$) \\ \hline
    \multirow{2}{*}{$\rho=0.5$} & Not spatial & 0.8650 (0.0001) & 0.1059 ($<10^{-4}$) & 0.6166 (0.0001) & 0.9686 ($<10^{-4}$)\\
    & Spatial & 0.9991 ($<10^{-4}$) & 0.0072 ($<10^{-4}$) & 0.5553 (0.0002) & 0.9999 ($<10^{-4}$) \\ \hline
    \multirow{2}{*}{$\rho=0.9$} & Not spatial & 0.2382 (0.0027) & 0.3057 (0.0002) & 0.8307 (0.0023) & 0.7920 (0.0004)  \\
    & Spatial & 0.9996 ($<10^{-4}$) & 0.0059 ($<10^{-4}$) & 0.3312 (0.0008) & 0.9999 ($<10^{-4}$) \\
    \hline
  \end{tabular}
\end{table}

\end{document}